\documentclass[a4paper,12pt]{article}
\usepackage[english]{babel}
\usepackage[utf8]{inputenc}
\usepackage{graphicx} 
\usepackage{fancyhdr} 
\usepackage{amsmath}
\usepackage{slashed}
\usepackage{dingbat}
\usepackage{cite}
\usepackage{amssymb}
\usepackage{hyperref}

\usepackage{mattens}
\usepackage{eucal}
\usepackage[usenames, dvipsnames]{color}
\numberwithin{equation}{section}

\usepackage{cite}
\begin{document} 
\begin{titlepage} 
 
\begin{flushright}
UoA  \\  
\today
 \end{flushright} 
\begin{center} 
\vspace*{1.5cm} 
 
{\Large{\textbf {{Gravitational Waves From No-Scale Supergravity }}}}\\ 
 \vspace*{10mm} 
 
 {\bf Vassilis~C.~Spanos and Ioanna~D.~Stamou}
\vspace{.7cm} 

{\it National and Kapodistrian University of Athens, Department of Physics, \\
 Section of Nuclear and  Particle Physics,  GR--15784 Athens, Greece} \\

\end{center} 
\vspace{2.cm}

\begin{abstract}
In this paper  we  study  four concrete  models,  based on no-scale supergravity 
with   SU(2,1)/SU(2)\-$ \times$ U(1) symmetry.  
\textcolor{black}{We modify either the K\"ahler potential or the superpotential, which are related to the no-scale theory with  this  symmetry. }
In this  scenario, the  induced Gravitational Waves,  
are calculated to be  detectable  by the future space-based observations such as LISA, BBO and DECIGO. The  models under study  are interrelated, as they all yield   the Starobinsky effective-like scalar potential in the unmodified case.  We evaluate numerically the  scalar power spectrum  and the stochastic background of the Gravitational Waves, satisfying   the observational Planck cosmological constraints for inflation.
  \end{abstract}
\end{titlepage} 

\tableofcontents

\section{Introduction}
\label{intro}

The detection of the Gravitational Waves (GWs) is 	undeniably regarded as  a milestone in  cosmology. The  GWs emitted by a  binary black hole merges,  reported 
 by the LIGO and Virgo collaborations~\cite{Abbott:2016blz,Abbott:2017vtc,Abbott:2017gyy,Abbott:2017oio,Abbott:2016nmj},  draw attention to both experimental and theoretical researches  for  exploring  the stochastic background of GWs.  The signal of the GWs is expected to be detectable  by future space-based  GW interferometers such as  LISA, BBO and DECIGO\cite{LISA:2017pwj, Yagi:2011wg}.

\textcolor{black}{The detection of these events  by  the  LIGO/Virgo  collaborations has been rekindled the old study of Primordial Black Holes (PBHs)\cite{10.1093/mnras/152.1.75,Hawking:1971tu,Carr:1974nx,Zeldovich:1967lct}.
Specifically, it has been claimed   that the PBHs, which are formed during radiation dominated epoch, can be  considered as reliable candidates for a large, or the whole amount of the Dark Matter (DM) of the Universe\cite{Sasaki:2018dmp,Garcia-Bellido:2017mdw,Ballesteros:2017fsr,Ozsoy:2018flq,Cicoli:2018asa,Gao:2018pvq,Dalianis:2018frf,Hertzberg:2017dkh,Mahbub:2019uhl,Nanopoulos:2020nnh,Stamou:2021qdk,Ezquiaga:2017fvi,Braglia:2020eai,Braglia:2020taf,Aldabergenov:2020bpt,Aldabergenov:2020yok}. 
In this case,  there is a  strong possibility to detect induced GWs' signals, which are generated during the radiation phase and  will be available in the future from  the aforementioned observations. The contribution of  the first order scalar perturbations, to the generation of second order perturbations associated to the GWs,    has already been studied in detail \cite{10.1143/PTP.37.831,matarese,Acquaviva:2002ud,Ananda:2006af,Baumann:2007zm,Espinosa:2018eve,Inomata:2019ivs}. }

Several mechanisms for producing enhancement in the scalar power spectrum have been proposed in the literature. In the case of the single field inflation,  the mechanism associated to the presence of an inflection point has been already studied 
extensively~\cite{Garcia-Bellido:2017mdw,Ballesteros:2017fsr,Ozsoy:2018flq,Cicoli:2018asa,Gao:2018pvq,Dalianis:2018frf,Hertzberg:2017dkh,Mahbub:2019uhl,Nanopoulos:2020nnh,Stamou:2021qdk,Ezquiaga:2017fvi,Aldabergenov:2020bpt,Aldabergenov:2020yok}. At the inflection  point,  both first and second derivatives of the effective scalar potential become approximately zero. This feature in the effective scalar potential,
significantly  decreases the velocity of the inflaton. 
This decrease  is imprinted in a peak in power spectrum. 
 The slow-roll approximation fails to give the correct results and one has to solve numerically 
 the exact equation of the  field perturbations,
 because  the slow-roll    parameter $\eta$  growths significantly. 
 In this work we will consider this mechanism for producing such amplifications. However, alternative  mechanisms have been   proposed regarding single field inflation. One of them is  models with a feature like  steps  in the potential\textcolor{black}{\cite{Adams:2001vc,Kefala:2020xsx,Dalianis:2021iig,Cai:2021zsp}.} Moreover, \textcolor{black}{models with ultra slow-roll inflation have been studied~\cite{Pattison:2021oen,Byrnes:2018txb}.} Two field inflationary model have also been proposed {\cite{Clesse:2015wea,Braglia:2020eai,Spanos:2021hpk,Gundhi:2020kzm,Pi:2021dft}.}

 The relation  between the first and second order perturbations, that is the scalar and the tensor power spectrum, 
gives the insight  to study the GWs by  analysing and measuring  only  the  scalar power spectrum \cite{10.1143/PTP.37.831,matarese,Acquaviva:2002ud,Ananda:2006af,Baumann:2007zm,Espinosa:2018eve,Domenech:2021ztg,Domenech:2019quo,Domenech:2020kqm}.    There are numerous  studies based on this scenario  \cite{Braglia:2020eai,Braglia:2020taf,Zhou:2020kkf,Gundhi:2020kzm,Fumagalli:2020nvq,Dalianis:2020cla,Domenech:2020ssp,Domenech:2020xin,Dalianis:2020gup,Xu:2019bdp,Gao:2021dfi,Dalianis:2021iig,Ballesteros:2020qam} and we briefly discuss some of them. In Refs.~\cite{Braglia:2020eai,Braglia:2020taf} the authors study  two scalar field models with a non-canonical kinetic term. Other two-field models, in order to generate  GWs have been studied in Refs.~\cite{Braglia:2020eai,Braglia:2020taf,Spanos:2021hpk,Gundhi:2020kzm,Zhou:2020kkf}. The production of GWs within a single field inflation has been considered in Refs.~\cite{Ballesteros:2020qam,Gao:2021dfi,Dalianis:2020cla,Dalianis:2021iig}.

 Supergravity (SUGRA) models can provide us with inflationary  potentials compatible with observational data. 
 To this end, a  significant enhancement in the   scalar power spectrum as described above, can occur~\cite{Nanopoulos:2020nnh,Stamou:2021qdk,Dalianis:2018frf,Gao:2018pvq,Aldabergenov:2020bpt,Aldabergenov:2020yok}.
In particular, it has been claimed that no-scale theory can be regarded as a theory worthy of consideration,  in order to study enhancement in power spectrum  \cite{Nanopoulos:2020nnh,Stamou:2021qdk}.  This theory treats many problematic issues such as the fine-tuning in order to obtain a vanishing cosmological constant  and it gives a natural solution to the $\eta$ problem \cite{Cremmer:1983bf,Ellis:1983sf,Ellis:1983ei,Ellis:1984bm,Lahanas:1986uc}.

In~\cite{Stamou:2021qdk} a mechanism to produce   PBHs was proposed, based on the breaking of the non-compact  SU(2,1)/SU(2)$ \times$ U(1) symmetry. 
It is proved that,    the effective scalar potentials related to inflation,
not only can  explain the production of PBHs, but also they
can conserve the transformation laws, related to the coset  SU(2,1)/SU(2)$ \times$ U(1).
In this work we will consider these potentials, in order to explain the generation of GWs and we will numerically evaluate the energy density. Moreover,  we generate GWs by modifying the kinetic term in the Lagrangian and  keeping the superpotentials, which lead to the Starobinsky-effective  potential such as the Wess-Zumino or the Cecotti, unchanged. All the models present in this work are in consistence with the Planck constraints of the spectral index $n_s$ and the  tensor-to-scalar ratio $r$ \cite{Ade:2015lrj,Akrami:2018odb}.  

The layout of the paper is  as follows: In  section 2 we briefly review  some basic aspects for the SUGRA theory and 
the  perturbation theory,  which are used in our analysis. In  section 3 we explain how the enhancement of power spectrum can be produced in the framework of no-scale theory.  In section 4 we present the GWs for our analysis.
  Finally, in section 5 we give our conclusions and perspectives.

\section{Basic Considerations}
\label{aspect}
In this section we discuss   two  basic considerations used in our analysis. 
The first  is an introduction to SUGRA theory, which is relevant to the inflation models 
we will employ below. 
The second is about  the general tools used 
for studying the field perturbations  during inflation. 

\subsection{General Aspect Of Supergravity}
In this subsection we present some basic aspects of SUGRA and particularly the    no-scale theory. 
The SUGRA Lagrangian coupled to matter is given in the following expression
\begin{equation}
\mathcal{L}=\frac{1}{2}R -K^{\Phi \bar{\Phi}} \partial_{\mu} \Phi \partial ^{\mu}\bar{\Phi}- V(\Phi, \bar{\Phi}) \, , 
\label{eq1}
\end{equation}
\noindent
where $K$ denotes the K\"ahler potential,  $W$ is the superpotential and the indices $\Phi$ and  $\bar{\Phi}$ denote the derivatives in respect to the corresponding  fields. 
We remark that throughout this study we work in reduced Planck units ($M_P=1$). 

The F-term of scalar potential is given as follows
\begin{equation}
V= e^{K} (D_{\Phi}WK^{\bar{\Phi} \Phi} D_{\bar{\Phi}}\bar{W} - 3|W|^2) \, , 
\label{eq2}
\end{equation}
 where
\noindent
\begin{equation}
D_{\Phi}W=\frac{\partial W}{\partial \Phi}+ \frac{\partial K}{\partial \Phi}W 
\end{equation}
is the K\"ahler convariant derivative. 

The minimal no-scale model is given by the K\"ahler potential in the present of a single chiral field $\Phi$ \cite{Cremmer:1983bf,Ellis:1983sf,Ellis:1983ei,Ellis:1984bm,Lahanas:1986uc} 
\begin{equation}
K=-3 \ln(\Phi+ \bar{\Phi}).
\label{eq3}
\end{equation}
By considering this K\"ahler potential, one can notice that the term $-3|W|^2$ in Eq.(\ref{eq2}) is vanishing because of the following expression \cite{Cremmer:1983bf}
\begin{equation}
K^{\Phi\bar{\Phi}}K_{\Phi}K_{\bar{\Phi}}=3.
\label{eq4}
\end{equation}

In the case of non-compact symmetry SU(2,1)/SU(2)$ \times$ U(1),  a no-scale supergravity model
is described  by the following two equivalent K\"ahler potentials \cite{Ellis:2018zya}
\begin{equation}
K=-3 \ln\left(T +\bar{T}- \frac{|\varphi|^2}{3}\right)
\label{eq5}
\end{equation}
and
\begin{equation}
K=-3\ln\left(1-\frac{|y_1|^2}{3}-\frac{|y_2|^2}{3}\right) 
\label{eq5a}
\end{equation}
where $y_1$, $y_2$, $T$ and $\varphi$ are chiral fields. The K\"ahler potential in $(y_1,y_2)$ basis is related to this  in
the $(T,\varphi)$ basis, as  analyzed in~\cite{Ellis:2013nxa}. The relevant  transformation is
\begin{equation}
 y_1=\Big(\frac{2\varphi}{1+2T}\Big) ,\quad y_2=\sqrt{3}\Big(\frac{1-2T}{1+2T}\Big) \, , 
 \label{eq2.3}
 \end{equation}
while  its  inverse is
  \begin{equation}
 T=\frac{1}{2}\Big(\frac{1 -y_2 / \sqrt{3}}{1 +y_2 / \sqrt{3}}\Big) ,\quad \varphi=\Big(\frac{y_1}{1 +y_2 / \sqrt{3}}\Big) \, .
 \label{k2(3)}
 \end{equation}
For a given superpotential,  in the $(y_1,y_2)$ basis, we can derive the corresponding in $(T,\varphi)$ 
by using the following relation~\cite{Ellis:2013nxa}
 \begin{equation}
 W(T,\varphi) \rightarrow \bar{W}(y_1,y_2)= \left( 1+y_2 /\sqrt{3} \right)^3W \, .
 \label{eq2.4}
 \end{equation}

\textcolor{black}{In order to present inflationary models which are in agreement  with the current observational data  on  CMB anisotropies~\cite{Ade:2015lrj,Akrami:2018odb}, we consider the Starobinsky-type potential. We note that in \cite{Ade:2015lrj,Akrami:2018odb}  the $\mathcal{R}^2$  inflationary models are in accordance with the current data.  Interestingly, the no-scale theory described above, provides us with potentials of this kind (Starobinsky-like potential) \cite{Cecotti:1987sa,Ellis:2013xoa}.  Furthermore,  there are other supergravity models,   which can lead to Starobinsky-like potential as well. We briefly mention the $\alpha$-attractor models,  which are described in \cite{Kallosh:2013hoa,Ferrara:2013rsa,Kallosh:2014rga}. In these models the Eq. (\ref{eq5a}) is multiplied by a term $\alpha$, which can be different to one:
\begin{equation}
K=-3\alpha \ln\left(1-\frac{|y_1|^2}{3}-\frac{|y_2|^2}{3}\right) .
\label{eq:aatractor}
\end{equation}
As a result the condition (\ref{eq4}) is not conserved. However, these models have other interesting features, e.g.   they are invariant  under the conformal transformations~\cite{Kallosh:2013hoa,Ferrara:2013rsa,Kallosh:2014rga}.}

\textcolor{black}{In our analysis}, the Starobinsky-like effective scalar potential can be derived with \textcolor{black}{two}  ways \textcolor{black}{from no-scale theory},  as it  is
described below. 
One is   this of the   Wess-Zumino model~\cite{Ellis:2013xoa}
\begin{equation}
W= \frac{\hat{\mu}}{2} \varphi^2 -\frac{\lambda}{3}\varphi^3 \, , 
\label{eq6}
\end{equation}
where $\hat{\mu}$ is a mass term and $\lambda$ a trilinear coupling in the Lagrangian.
This is written in the  $(T,\varphi)$ basis. 
In order to get  the Starobinsky-like effective scalar potential, 
we use the Eq.(\ref{eq2}) and
we assume that  the inflationary direction is   ${\textrm{Im}}\varphi={\textrm{Im}} T=0$ and ${\textrm{Re}}T=c$. 
\textcolor{black}{We  define   $\hat \mu = \mu \sqrt{c/3}$. }
In addition,  in order to evaluate the effective  potential we need to fix the
non-canonical kinetic term. To do this  we use the following redefinition of the field
\textcolor{black}{\begin{equation}
 \varphi= \sqrt{3} \tanh\left( \frac{\phi}{\sqrt{6}}\right)
\label{eq8}
\end{equation}}
\textcolor{black}{where $\phi$ is the field with fixed the non-canonical kinetic term. } \textcolor{black}{From Eq.(\ref{eq2}) we get 
\begin{equation}
V=\mu^2 e^{-\sqrt{2/3}x}\sinh^2\left( \frac{\phi}{\sqrt{6}}\right),
\label{eq:pot:eno}
\end{equation}
where we have assumed $\lambda/ \mu=1 /3$. }

The superpotential in  Eq.(\ref{eq6}) can be written in $(y_1,y_2)$ basis, using  the Eqs.~(\ref{k2(3)}) and (\ref{eq2.4})~\cite{Ellis:2013nxa}. Hence,  we get 
\begin{equation}
W=\frac{\hat\mu}{2} \Big(y_1^2 +\frac{y_1^2 y_2}{\sqrt{3}}\Big) -\lambda \frac{y_1^3}{3} 
\label{eq7}
\end{equation}
and the kinetic term is fixed using the  redefinition
\textcolor{black}{\begin{equation}
y_1= \sqrt{3} \tanh\left(\frac{\phi}{\sqrt{6}}\right) \, ,
\label{eq13}
\end{equation}}
where the Starobinsky effective scalar potential appears 
assuming that the   $y_1$ is the inflaton field and  $y_2$ is  the modulo. \textcolor{black}{ The resulting effective scalar potential is the same with this  derived in the $(T, \varphi)$ basis.}

Alternatively, the  Starobinsky-like  potential can be derived  by the Cecotti superpotential as 
given in $(T,\varphi)$ basis as \cite{Cecotti:1987sa}
\begin{equation}
W=\sqrt{3} \, m \, \varphi \, \left(T-\frac{1}{2}\right) \, ,
\label{eq9}
\end{equation}
\textcolor{black}{where $m$ is a mass term. The} Starobinsky effective scalar potential arises thought the direction $\varphi= \text{Im}T=0$. In order to have canonical kinetic terms, one has to use  the field  redefinition 
\begin{equation}
T=\frac{1}{2}e^{\sqrt{\frac{2}{3}} \phi}\,  .
\label{eq10}
\end{equation}

\textcolor{black}{In general, the kinetic term can be fixed by the  transformation 
\begin{equation}
\label{eq12}
\frac{1}{2}\partial_{\mu}\phi \partial^{\mu}\phi= K_{\chi \chi}\partial_{\mu}x \partial^{\mu}\chi\Rightarrow\frac{d \phi}{d \chi}= \sqrt{2 K_{\chi \chi }} \, .
\end{equation}
Although the relation ~(\ref{eq12})  
in the case of the K\"ahler potentials~(\ref{eq5}) and (\ref{eq5a}) 
yields  analytical solution, it is also generally useful even in the cases that  there is no analytical solution. 
In the following, we will consider both analytical and numerical solutions of  Eq.~(\ref{eq12}).}

The superpotential   in  Eq.~(\ref{eq9}) can be written in the $(y_1,y_2)$ basis as
\begin{equation}
W=m\Big( -y_1y_2 +\frac{y_2y_1^2}{\sqrt{3}}\Big )
\label{eq11}
\end{equation}
and as before the redefinition for the  canonical kinetic term is  
\textcolor{black}{\begin{equation}
y_1= \sqrt{3} \tanh\left(\frac{\phi}{\sqrt{6}}\right) \, ,
\label{eq13b}
\end{equation}}
with  $y_1$ being the inflaton field
\textcolor{black}{ and $y_2$ the modulo. If we evaluate the effective potential,  we can obtain the same Starobinsky-like potential in both  bases  $(T,\varphi)$ and $(y_1,y_2)$. }

\textcolor{black}{Therefore, there is an important equivalence between the models of Wess-Zumino and Cecotti,  as they can provide us with Starobinsky-like potential, when they are embedded in no-scale theory \cite{Ellis:2018zya}. Moreover, when we study the forms written in $(y_1,y_2)$ basis in Eq.(\ref{eq5a}) and the corresponding superpotential given in (\ref{eq7}) and  (\ref{eq11}) we can obtain that  these models obey the  symmetry  identified by the matrix
\begin{gather}
U
 =
  \begin{bmatrix}
   \alpha &
  \beta & 0 \\
   -\beta^* & \alpha^* & 0 \\
   0& 0& 1 
   \end{bmatrix} \, , 
\end{gather}
\noindent
where $\alpha,\beta \in\mathbb{C} $ and $|\alpha|^2+|\beta|^2=1$. By this matrix and the analysis shown in Ref.~\cite{Ellis:2018zya} we can obtain the following transformation laws for the fields
\begin{equation}
y_1 \rightarrow \alpha y_1 + \beta y_2, \quad y_2 \rightarrow - \beta^* y_1 +\alpha ^* y_2.
\label{eq2.7}
\end{equation}
This symmetry is analyzed in \cite{Ellis:2018zya}.  In the next section we will present models that satisfies  this symmetry. }


\subsection{Curvature Perturbations}
\label{Curvature Perturbations}

The equation of motion of the inflaton field is  
\begin{equation}
\label{eq21}
\phi'' +3\phi'- \frac{1}{2} \phi'^3 +\left(3- \frac{1}{2} \phi'^2\right) \frac{d\ln V(\phi)}{d \phi}=0 \, , 
\end{equation}  
where  primes  denote the derivative in efold time. 

The general  Friedmann-Robertson-Walker (FRW) metric written in the conformal Newtonian gauge is \cite{MUKHANOV1992203,Mukhanov:1988jd}
\begin{equation}
ds^2 = - a^2 \left( 1+2 \Phi \right) d \tau^2 +a \left(  (1+2 \Psi)\delta_{ij} +\frac{1}{2}h_{ij}\right)dx^idx^j \, ,
\label{eq1.1}
\end{equation}
where  $\tau$ is the conformal time,  $a$ is the scale factor, $h_{ij}$ are tensor perturbations, $\Phi$ and $\Psi$ are the Bardeen potentials, which are equal   in the absence of anisotropy in the stress energy tensor. 
Assuming that  the perturbation of the field is  $\phi+ \delta \phi$, the equation of the curvature perturbation takes the form \cite{Ringeval:2007am}
\begin{equation}
 \label{eq1.2} 
{\delta \phi''=- \left(  3-\frac{1}{2} \phi'^2 \right) \delta \phi'-\frac{1}{H^2}\frac{d^2V}{d\phi^2} \delta \phi -\frac{k^2}{a^2 H^2} \delta \phi+4 \Psi' \phi'-\frac{2 \Psi}{H^2}\frac{dV}{d \phi}}  \, , 
\end{equation}
where the Bardeen potential $\Psi$ results from  the equation
\begin{equation}
 \label{eq1.3}
{\Psi''=-\left(  7-\frac{1}{2} \phi'^2 \right) \Psi'- \left( 2\frac{V}{H^2} +\frac{k^2}{a^2 H^2}\right)\Psi -\frac{1}{H^2}\frac{dV}{d\phi} \delta \phi\, .}
\end{equation} 
With $H$ we denote the Hubble parameter, which is  
\begin{equation}
 \label{eq1.4}
{H^2=\frac{V}{(3- \frac{1}{2} {\phi'}^2)} \, .}
\end{equation}
The initial condition for the perturbation,  assuming  Bunch-Davies vacuum, is
\begin{equation}
aQ \rightarrow \frac{e^{-ik\tau}}{\sqrt{2k}} \, , 
\end{equation}
where $Q$  is the Mukhanov-Sassaki variable. The complete expressions for the initial conditions for the perturbation,  as well as the initial conditions for the Bardeen potential are  
\begin{equation}   
\begin{split}
{\delta \phi_{ic}=\frac{1}{\sqrt{2 k}} \frac{1}{a_{ic}}, \quad\left( \delta  \phi '\right)_{ic}= -\frac{1}{a_{ic}\sqrt{2k}}\left(1+i\frac{k}{a_{ic}H_{ic}}\right)}
\label{26b}
\end{split}
\end{equation}
and
{\begin{equation}
\begin{split}
\Psi_{ic}=&\frac{1}{2 \left( \varepsilon_{H,ic} -\frac{k^2}{a_{ic}^2 H_{ic}^2} \right) } \left(   \phi'_{ic}  \delta\phi'_{ic}+\delta \phi_{ic}\left[3\left(\phi'\right)_{ic} +\frac{1}{H_{ic}^2}\left(\frac{dV}{d\phi}\right)_{ic}\right]\right)\\
 \Psi'_{ic}=& \frac{1}{2} \left( \phi'\right)_{ic} \delta \phi_{ic} -\Psi_{ic}\, .
\end{split}
\end{equation}}
With the indices $ic$ we denote the initial conditions and with the $\varepsilon_H$ we denote the slow roll parameter
\begin{equation*}
  \varepsilon_H=\frac{1}{2} \phi'^2. 
\end{equation*}

We note that these equations are valid for a Lagrangian with a canonical kinetic term. Otherwise one should  take into account the field transformation in order to fix the canonical kinetic term, as we have mentioned in the previous subsection in Eq.(\ref{eq12}). 

Using Eqs.~(\ref{eq21}) -(\ref{eq1.4}) we can evaluate the power spectrum as
\begin{equation}
\label{eq1.5}
P_R=\frac{k^3}{2 \pi^2} \left|R_k \right|^2,
\end{equation}
where $k$ is the comoving wavenumber of the Fourier mode and
{\begin{equation}
R_k=\Psi +\frac{\delta \phi }{\phi'} \, .
\label{eq1.6}\end{equation}}
The numerical procedure,  we follow is descibed  in Refs. \cite{Nanopoulos:2020nnh,Stamou:2021qdk}.

\textcolor{black} {There are many works where the analytical evaluation of enhanced power spectrum has been studied, such as \cite{Kefala:2020xsx,Dalianis:2021iig,Byrnes:2018txb}.  The power spectrum in the vicinity  of 
the peak  can be approximated analytically by the power-law spectrum which is given as \cite{Kohri:2018awv,Atal:2018neu}:
\begin{equation}
P_R= \begin{cases}
      0 & \text{if $k<k_{peak}$}\\
     P_0 \left( \frac{k}{k_{peak}}\right)^{-n} & \text{if $k>k_{peak}$} \, \,  ,
\end{cases}  
\label{eq:pr_analu}
\end{equation}
 where $k_{peak}$ is the comoving wavenumber  of the peak.  $P_0$ is related to the peak amplitude  and  $n$ is related to  the spectral index.  In the following we present  both numerical and analytical results for the sake of  comparison. }
 

\textcolor{black}{The models we  study in this work are based on $\mathcal{R}^2$ inflation and  are consistent with the current observational data of inflation. 
To this end, we use the slow-roll parameter  $\eta_H$, defined as
\begin{equation}
\eta_H= \varepsilon_H- \frac{1}{2} \frac{d \ln   \varepsilon_H}{dN} \,. 
\end{equation}
The slow-roll parameters $\varepsilon_V$ and $\eta_V$ are  defined also  in terms of the inflationary potential as 
\begin{equation}
\varepsilon_V= \frac{1}{2}\left(\frac{V'(\phi)}{V(\phi)} \right)^2 , \quad \eta_V=\frac{V''(\phi)}{V(\phi)}\, , 
\end{equation}
where primes  denote the derivatives in  respect of the field $\phi$.}
\textcolor{black}{The spectral index is defined as 
\begin{equation}
n_s-1 =\frac{d\ln P_R}{d \ln k}  \,. 
\end{equation}
In terms of slow roll conditions for the prediction of the observable  $n_s$, we get 
\begin{equation}
\label{ns_def}
n_s \simeq 1+2 \eta_V -6\varepsilon_V \, . 
\end{equation} 
The prediction for the ratio tensor-to-scalar $r$ are
\begin{equation}
\label{r_def}
 r \simeq 16 \varepsilon_V.
\end{equation}  
The corresponding values for these quantities, which are evaluated at the pivot scale of $k_*=0.05Mpc^{-1}$ are   \cite{Akrami:2018odb} 
\begin{equation}
  n_s =
    \begin{cases}
      0.9659 \pm 0.0041 & \text{Planck TT,TE,EE+lowEB+lensing}\\
      0.9651 \pm 0.0041 & \text{Planck TT,TE,EE+lowEB+lensing+BK15}\\
       0.9668 \pm 0.0037 &  \text{Planck TT,TE,EE+lowEB+lensing+BK15+BAO}
    \end{cases}       
    \label{ns-obs}
\end{equation}
and
\begin{equation}
r <
    \begin{cases}
      0.11 & \text{Planck TT,TE,EE+lowEB+lensing}\\
      0.061  & \text{Planck TT,TE,EE+lowEB+lensing+BK15}\\
       0.063  & \text{Planck TT,TE,EE+lowEB+lensing+BK15+BAO} \, .
    \end{cases}  
    \label{ratio-obs}
\end{equation}
In the following we present    the predictions 
for    the observables of   the proposed models and we 
will check their consistency against the  Planck  data.
}


\section{Gravitational Waves from the  No-Scale models}
\label{gwwbns}

Below we will study how modifications  in the superpotential or in the K\"ahler potential can be used in order to explain the generation of GWs.  We present two different ways to produce such peaks in power spectrum, which as we will see it is related to the spectrum of GWs. The first is by modifying the Wess-Zumino   and Cecotti superpotential, given in Eqs.~(\ref{eq6}), ~(\ref{eq7}),~(\ref{eq8}) and ~(\ref{eq9}).  The second is by modifying the K\"ahler potential, meaning the kinetic term on the Lagrangian, taking into account that both Wess-Zumino superpotential and Cecotti superpotential are unchanged. Therefore, we present two different ways in order to get  significant peaks in the power spectrum by \textcolor{black}{conserving (modifying superpotentials)} and breaking \textcolor{black}{(modifying K\"ahler potential)} the SU(2,1)/SU(2)$ \times$ U(1) symmetry. All models give the Starobinsky effective scalar potential in the unmodified case. 

\subsection{Modifying the superpotential}
  \label{attr}
It was shown that the non-compact  SU(2,1)/SU(2)$ \times$ U(1) symmetry leads to the same  Starobinsky-like effective scalar potential, demonstrating an important equivalence between  the different models \cite{Ellis:2018zya}. Moreover, it can be shown that this symmetry can be preserved \cite{Stamou:2021qdk}  under proper modification of the superpotential, in order to generate a  peak in power spectrum and produce significant amount of the   PBHs in the Universe. Therefore, we consider the no-scale supergravity coset  SU(2,1)/SU(2)$ \times$ U(1). This coset can be identified by two equivalent forms of the  K\"ahler potential
\begin{equation}
\begin{split}
K=-3\ln(1-\frac{|y_1|^2}{3}-\frac{|y_2|^2}{3}) \\
 K=-3 \ln(T +\bar{T}- \frac{|\varphi|^2}{3})\, , 
\end{split}
\label{eq2.2}
\end{equation}
which can be written in two basis, the $(T,\varphi)$ and $(y_1,y_2)$.

First, by a proper modification of the Wess-Zumino model, we can derive the same effective scalar potential in both $(T, \varphi)$ and $(y_1,y_2)$ basis. The superpotential
 adopted in Ref. \cite{Stamou:2021qdk}, for modified the Wess-Zumino model  in $(y_1,y_2)$ basis, is
\begin{equation}
W=\left[ \frac{\hat\mu}{2} \left(y_1^2 +\frac{y_2^2 y_1}{\sqrt{3}}\right) -\lambda \frac{y_1^3}{3}\right]  
\left[ 1+e^{-b_1y_1^2} \left(c_1 {y_1}^2+c_2{y_1}^4 \right)\right] \, , 
 \label{eq2.5}
\end{equation}
where $\hat\mu$, $\lambda$, $b_1$, $c_1$ and $c_2$ are free parameters, fixed by observations.  The effective scalar potential can be calculated by using  Eq.~(\ref{eq2}) and   the redefinition of the field, which is given in Eq.~(\ref{eq13}). Using this modification the resulting effective scalar potential is not affected by the transformations   $y_1 \rightarrow -y_1$ and $y_2 \rightarrow -y_2$. Alternatively, using   $y_1 \rightarrow -y_2$ and $y_2 \rightarrow y_1$, one can derive the same effective scalar potential, in the inflationary direction \textcolor{black}{ $y_2={\textrm{Im}} \, y_1=0$, by fixing the non-canonical kinetic term as }
\textcolor{black}{\begin{equation}
y_1=-\sqrt{3}\tanh\left( \frac{\phi}{\sqrt{6}}\right) \,. 
\end{equation}}
This analysis is presented in~\cite{Ellis:2018zya}.

\begin{figure}[h!]
\centering
\includegraphics[width=80mm]{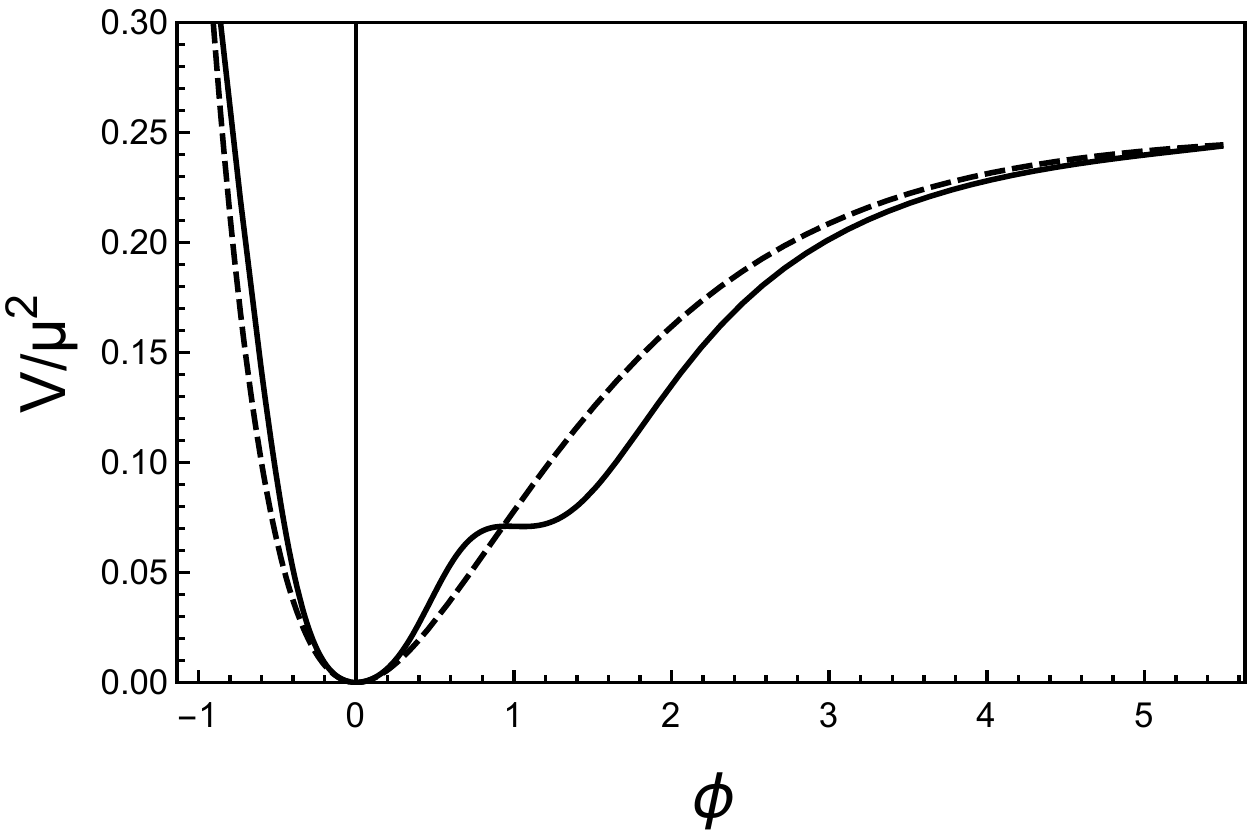}

\caption{\textcolor{black}{The effective potential (solid line) for the superpotential in
Eq.~(\ref{eq2.5})  with the choice of parameters $\lambda / \mu=1/3$, $c_1=1.9$, $c_2=-0.73$ and $b_1=3.9615$. The dashed like  corresponds to the Starobinsky potential. } }
\label{fpot}
\end{figure}

 The superpotential in Eq.~(\ref{eq2.5}) can be written in the $(T,\varphi)$ basis, if one considers  the Eq.~(\ref{eq2.4}). Hence 

\begin{equation}
 W=\left( \frac{\hat{\mu}}{2}\varphi^2 -\frac{\lambda}{3} \varphi^3\right)\left(  1+ e^{-b_1 \left( \frac{2 \varphi}{1+ 2 T} \right)^2} \left[ c_1 \left( \frac{2 \varphi}{1+ 2 T} \right)^2+ c_2 \left( \frac{2 \varphi}{1+ 2 T} \right)^4\right] \right) \, , 
\label{eq2.5b}
\end{equation}
where $\varphi$ is the inflaton field and $T$ is the modulo. This yields  the same effective scalar potential as in Eq.~(\ref{eq2.5}) using  the redefinition of the field (\ref{eq8}), in order to get  canonical kinetic terms. \textcolor{black}{ In Fig. \ref{fpot} we present the effective scalar potential given for the superpotential  (\ref{eq2.5}) (or equivalently Eq.(\ref{eq2.5b})). The details for this calculation  are described in Appendix \ref{sect:app1}. In order to derive this potential we redefine the field in order to have canonical kinetic terms. We notice that there is a region  around $\phi \approx 1.4$,  where there is a near inflection point. We expect that this feature can be imprinted in the scalar power spectrum. On the other hand, there are many mechanisms in order to obtain an enhancement in scalar power spectrum, such as two field models,  models with bumpy features\cite{Mishra:2019pzq,Zheng:2021vda}. In Appendix \ref{sect:app2} there is a comparison between some models.   In  Fig. \ref{fpot}  we also depict the case of $c_1=c_2=0$ which correspond to the  Starobinsky potential. }

In addition to that, by a proper modification in Cecotti superpotential, given in (\ref{eq9}), we can produce significant peak in the power spectrum in the  $(y_1,y_2)$ basis. In Ref.\cite{Stamou:2021qdk}  the following superpotential   is considered

\begin{equation}
 W=m\left(-y_1y_2 
  +\frac{y_2y_1^2}{l\sqrt{3}}\right)\left(1+c_3e^{-b_2{y_1}^2}{y_1}^2\right) \, ,
 \label{eq2.6}
\end{equation}
where $m$, $l$, $c_3$, $b_1$ are the parameters of the model. 
This choice of modification is made for keeping the resulting effective scalar potential unchanged under the 
transformations  $y_1 \rightarrow -y_1$, $y_2 \rightarrow -y_2$. 
Also, if one considers the case of exchanging the modulo and the inflaton field by changing
 $y_1 \rightarrow -y_2$ and $y_2 \rightarrow y_1$, the effective scalar potential remains the same. \textcolor{black}{For more detail see Refs.~\cite{Stamou:2021qdk}.}
 
In the $(T,\varphi)$ basis, the superpotential (\ref{eq2.6}) takes the following form using  Eq.~(\ref{eq2.4})
\begin{equation}
W= \frac{\sqrt{3}}{2} m \varphi \Big( \frac{1}{2} -T\Big) \left( -1-2T -\frac{1-2T}{l} \right) \left[ 1+ 3c_3 e^{-3b_2 \left( \frac{1-2T}{1+2T} \right)^2}\left( \frac{1-2T}{1+2T} \right)^2\right] \, .
 \label{eq2.6b}
\end{equation}
This gives the same effective scalar potential as in Eq.~(\ref{eq2.6}), if we consider the redefinition of the field of Eq.~(\ref{eq10}).

The superpotentials (\ref{eq2.5}) and (\ref{eq2.6}) not only preserve the same resulting effective scalar potential in both $(y_1,y_2)$ and $(T, \varphi)$ basis,  by the transformation given in Eq. (\ref{eq2.4}), but also they can produce significant enhancement in the power spectra, explaining the production of PBHs, as shown in \cite{Stamou:2021qdk}. In the following we will show that these peaks can also produce   GWs spectra sizeable enough to be detected in the 
corresponding experiments. These peaks are related to the inflection points in effective scalar potential, which one can find by a proper choice of parameters, as shown in \cite{Stamou:2021qdk}.  

\textcolor{black}{To sum up, the new terms that are present in  Eqs.(\ref{eq2.5}) and (\ref{eq2.6})  not only  preserve the transformation laws of the symmetry SU(2,1)/SU(2)$ \times$ U(1) as given in Eq.~(\ref{eq2.7}), but also  result to a significant  enhancement in scalar power spectrum. Hence, these  modifications have the following interesting properties.  First,  they conserve  the SU(2,1)/SU(2)$ \times$ U(1) symmetry. Second, they give the same form of the potential in both  bases  $(y_1,y_2)$ and $(T,\varphi)$. Third,  they  satisfy   the no-scale condition (\ref{eq4}), in order to ensure the  vanishing of the cosmological constant. Last but not least, they can result to  an inflection point under proper choice of parameters and finally, therefore  can  yield to a significant amount  of PBHs and GWs. }

%

\subsection{Modifying the K\"ahler potential}
  \label{attr}
In the previous section we study the generation of significant peaks in power spectrum by modifying well-known superpotentials, which yield to the Starobinsky-like effective scalar potential. These significant peaks   can be also presented by modified only the K\"ahler potential and keeping the superpotential unchanged.  Hence, this amplification can be found via   breaking the  SU(2,1)/SU(2)$ \times$ U(1) symmetry by adding extra terms in the kinetic part of the Lagrangian. In this section we present two models in order to produce such peaks, one   by keeping the Wess-Zumino superpotential (\ref{eq6}) unchanged and the other by keeping the Cecotti superpotential unchanged (\ref{eq9}).

In a previous work it was shown that sizable amplification in the power spectrum 
can be produced through the scheme\cite{Nanopoulos:2020nnh}
\begin{equation}
K=-3 \ln( T+ \bar{T} - \frac{\varphi \bar{\varphi}}{3}+ c_3e^{-b_3(\varphi +\bar{\varphi})^2}(\varphi +\bar{\varphi})^4) \,,
\label{eqwess}
\end{equation} 

\begin{equation}
W= \frac{\hat \mu}{2} \varphi^2  - \frac{\lambda}{3} \varphi^3 \, , 
\label{equation1}
\end{equation} 
where $c_3$ and $b_3$ are free parameters of the model. This model has to deal with just an extra parameter $b_3$, which gives us both  peaks in power spectrum and the correct prediction of spectral index $n_s$ and the tensor-to-scalar ratio $r$. The advantage of this model is that it leads to a reduced level of fine-tuning, as we will see later.

Also it is considered that the superpotential is given by the Cecotti form, Eq.~ (\ref{eq9}). Specifically, we have 
\begin{equation}
K=-3 \ln( T+ \bar{T} - \frac{\varphi \bar{\varphi}}{3}+ F( T+ \bar{T},\varphi+ \bar{\varphi}))
\label{eq2.8}
\end{equation} 

\begin{equation}
W=\sqrt{3}m \varphi\left(T- \frac{1}{2}\right)
\label{eq2.9}
\end{equation}
\noindent
where $F( T+ \bar{T},\varphi+ \bar{\varphi})$ is a function of both chiral fields. Choosing
\begin{equation}
\label{eq2.10}
 F( T+ \bar{T},\varphi+ \bar{\varphi})=e^{-d (T+ \bar{T})}\left(c_4 e^{-b_4(T+ \bar{T})}(T+ \bar{T})^2 +\bar{\lambda}(\varphi+ \bar{\varphi}+ \bar{\mu})\right) \, ,
\end{equation}
\noindent
where $b_4$, $c_4$, $\bar{\lambda}$, $\bar{\mu}$  and $d$ are free parameters. Although,  this scheme is more complicated than the previous one, one can derive effective scalar potential which is flat in the large values of field.

To summarize, we have presented  two different mechanisms,  in order to explain the production of GWs. The first of them is by modifying the superpotential and preserve the transformation laws of the symmetry  SU(2,1)/SU(2)$ \times$ U(1), which provide us with an important equivalent between the models  \cite{Ellis:2018zya}. The other   is by modifying the K\"ahler potential and keeping the superpotential unchanged. In all models presented in this work we remark that we are in complete consistence with the Planck constraints.

\section{Producing Gravitational Waves}
\label{pgw}

Based on the previous analysis  we can  evaluate the amount of GWs produced  during the radiation dominated epoch. The GWs are calculated by the tensor perturbations (\ref{eq1.1}). The equation of motion of GWs  reads as
\begin{equation}
\frac{d^2h_{k}}{d \tau}+2aH\frac{dh_{k}}{d \tau}+k^2 =  \hat{\mathcal{S}}_k\, ,
\label{eq4.51}
\end{equation}
where $h_{k}$ corresponds to the tensor metric perturbation written in Fourier space from Eq.~(\ref{eq1.1}) and the $ \hat{\mathcal{S}_k}$ is the source term  in Fourier space. In the radiation dominated era the solution of Eq.(\ref{eq4.51}) is obtained through the method of Green function as
\begin{equation}
h_k(\tau)=\frac{1}{a(\tau)} \int d \tilde{\tau} \, G_k(\tau, \tilde{\tau}) \, a(\tau)\,  \hat{\mathcal{S}}_k \, , 
\label{eq4.52}
\end{equation}
where the Green function is given as follows
\begin{equation}
G_k(\tau, \tilde{\tau})=\frac{\sin\left( k(\tau- \tilde{\tau}) \right)}{k} \,  \Theta(\tau-  \tilde{\tau}) \, .
\label{eq4.53}
\end{equation}

The power spectrum of GWs is related to the scalar power spectrum, as the $h_k$ can be expressed 
in curvature perturbation $R$ as
\begin{equation}
h_k(\tau)=\frac{4}{9} \int \frac{d^3 p}{(2 \pi)^3} 
\frac{1}{k^3 \tau}e(\textbf{k},\textbf{p}) R(\textbf{p})R(\textbf{k}-\textbf{p})
\, \left[ I_c(\tilde{x},\tilde{y})\cos(k\tau) + I_s(\tilde{x},\tilde{y})\sin(k\tau) \right] \, ,
\label{eq4.54}
\end{equation}
where $\tilde{x}=p/k$, $\tilde{y}=|\textbf{k}-\textbf{p} |/k$ and  $e$ is the polarization tensor of the graviton. 
The functions  $I_c$ and $I_s$  are given below.
Hence, the spectrum of second order GWs can be evaluated through the first order scalar perturbations \cite{Acquaviva:2002ud,Ananda:2006af,Baumann:2007zm,Espinosa:2018eve}.

 The general expression, in order to generate the present-day  GWs density function ${\Omega}_{GW}$, is given by \cite{Maggiore:1999vm}
 \begin{equation}
 {\Omega_{GW}(k)}=\frac{1}{24}\left( \frac{k}{a H}\right)^2 \overline{P_h(\tau,k)} \, ,
 \label{eq4.55}
 \end{equation}
 where $P_h$ is the tensor power spectrum and the overline denotes the average over time.
The $P_h$  is directly related to the  scalar power spectrum. Moreover, 
the energy density of the GWs in terms of scalar power spectrum is~\cite{Espinosa:2018eve,Kohri:2018awv,Inomata:2018epa}
\begin{equation}
{\Omega_{GW}(k)}=\frac{\Omega_r}{36} \int^{\frac{1}{\sqrt{3}}}_{0}\mathrm{d} d
\int ^{\infty}_{\frac{1}{\sqrt{3}}}\mathrm{d} s \left[  \frac{(s^2-1/3)(d^2-1/3)}{s^2+d^2}\right]^2\, 
P_{R}(kx) \,  P_{R}(ky) \,(I_c^2+I_s^2).
\label{eq4.1}
\end{equation}
The radiation density $\Omega_r$ has the present day value $\Omega_r \approx 5.4 \times 10^{-5}$. The variables $x$ and $y$ are
\begin{equation}
x= \frac{\sqrt{3}}{2}(s+d), \quad  y=\frac{\sqrt{3}}{2}(s-d).
\label{eq4.2}
\end{equation}
Finally, the functions $I_c$ and $I_s$ are given by the expressions 
\begin{gather}
I_c=-36 \pi \frac{(s^2+d^2-2)^2}{(s^2-d^2)^3}\Theta(s-1)\\
I_s=-36 \frac{(s^2+d^2-2)^2}{(s^2-d^2)^2}\left [ \frac{(s^2+d^2-2)}{(s^2-d^2)} \log\left| \frac{d^2-1}{s^2-1} \right| +2\right].
\label{eq4.3}
\end{gather}
Using that  
\begin{center}
$1 \, {\rm Mpc}^{-1}= 0.97154\times \, 10^{-14} \,  {\rm s}^{-1}$ and $k=2\pi f$
\end{center}
we can evaluate the energy density of power spectrum as a function of the frequency $f$. \textcolor{black}{We notice  that we must consider the possible  modification  of  the relativistic degrees of freedom, as  described in~\cite{Inomata:2018epa}. 
Although a detailed discussion on this,  goes beyond the scope of this work, 
  we have checked  numerically that our  results have a quite weak dependence on this modification. }

\textcolor{black}{An analytical estimation of the energy density of the GWs has been shown in  \cite{Kohri:2018awv}. 
The power spectrum can be approximated by  the  power-law spectrum as in  Eq.(\ref{eq:pr_analu}). In  Ref. \cite{Kohri:2018awv} the following approximation is considered
\begin{equation}
{\Omega_{GW}(k)}=Q(n) \left(P_0 \left( \frac{k}{k_{peak}}\right)^{-n} \right)^2,
\label{eq:gwUV}
\end{equation}
where the coefficients $Q(n)$ are given in   \cite{Kohri:2018awv} for specific values of $n$.  }
\textcolor{black}{This equation is valid  in ultraviolet (UV) region, if $k \gg k_{peak}$\cite{Atal:2021jyo,Liu:2020oqe}.  There are two assumption for the value of $n$
\begin{equation}
n =
    \begin{cases}
     n_{UV} &, \quad 0< n_{UV} <4\\
    2+ n_{UV}/2  &, \quad n_{UV} >4 \, , 
    \end{cases}
\end{equation} 
where $n_{UV} $ is a positive constant and it is referred as the spectral index in UV region.  For $k \ll k_{peak}$ in the infrared (IR) regions
we have the following approximation
\begin{equation}
{\Omega_{GW}(k)}=Q(n) P_0^2 \left( \frac{k}{k_{peak}}\right)^{3} \,  .
\end{equation}}

\textcolor{black}{ In \cite{Yuan:2019wwo,Cai:2019cdl} is shown that there is a general behaviour of the power spectrum of GWs in the  IR regions.
In particular, in the IR scales the energy density of GWs has been shown 
that follows the power law~\cite{Yuan:2019wwo}:
\begin{equation}
\Omega_{GW} \propto	f^{3-2/ \ln(f_c/f)},
\label{eq:gwIR}
\end{equation}
where $f_c$ is   roughly the peak of the frequency. 
In the region of the peak this approximation becomes
\begin{center}
 $\Omega_{GW}\propto	 f^{2-2/ \ln(f_c/f)}$, 
\end{center} 
and it  is valid   if the power spectrum of scalar curvature perturbation is narrow~\cite{Yuan:2019wwo}. 
}

\subsection{Modifying the superpotential}
  \label{attr}
Now we turn to  the results of the power spectrum and the stochastic spectrum of GWs in the case of superpotentials (\ref{eq2.5}) and (\ref{eq2.6}), where  the K\"ahler potential gets the form in $(y_1,y_2)$ basis  given in Eq.~({\ref{eq5a}}). By evaluating the effective scalar potential as shown in Eq.(\ref{eq1}), we can   get  inflection points, which 
can yield  significant enhancement in the  power spectrum,  as
calculated from  Eq.~(\ref{eq1.5}).  We remark that using the K\"ahler potential in the $(T,\varphi)$ basis given from Eq.~(\ref{eq2.2}) and the corresponding superpotential from Eqs.~(\ref{eq2.7}) and (\ref{eq2.9}),  we derive the same effective scalar potential and hence the same results. In the following, we will use the  basis $(y_1,y_2)$, in spite  the fact that we derive the same effective potential in the basis ($T,\varphi$) if we consider the transformation laws given in Eq.~(\ref{eq2.3}) and analyzed in \cite{Ellis:2018zya,Stamou:2021qdk}.

Having evaluated the effective scalar potential using  the superpotentials in (\ref{eq2.5}) and (\ref{eq2.6}), as well as   the K\"ahler potential  in ({\ref{eq5a}}), we proceed to the numerical integration of the background dynamics, as given in Eq.(\ref{eq21}), and  the  perturbation of the field as given in the Eq.~(\ref{eq1.2}) and  Eq.~(\ref{eq1.3}). The numerical procedure, which we follow, 
is shown in Refs.~ \cite{Nanopoulos:2020nnh,Stamou:2021qdk}. For initial conditions of the field we use those which  that  predict   $n_s$ and $r$ compatible to Planck measurements~\cite{Ade:2015lrj,Akrami:2018odb} 
and result to a  peak in power spectrum 
in the appropriate  place,
in order to explain the generation of GWs.  
The values of the initial condition, as well as the prediction for $n_s$ and $r$ for the cases (\ref{eq2.5}) and (\ref{eq2.6}) are given in Table~\ref{tabu1} respectively. With $\phi$ we denote the field after the redefinition given in Eq.~(\ref{eq13}) for obtaining  canonical kinetic terms.

\begin{table}[h!]
\centering
 \begin{tabular}{|c|c|c|c|}
\hline
 & $\phi_{ic}$ & $n_s$ & $r$\\
\hline
1 & $4.98$ &$0.9691$ &$0.0103$  \\
\hline
2 & $4.84$ &$0.9639$ &$0.0153$\\
\hline

\end{tabular}
 \caption{Initial conditions for the effective scalar potential in  Eqs. (\ref{eq2.5}) and (\ref{eq2.6}) and the predictions for $n_s$ and $r$.}
 \label{tabu1}
\end{table}

\begin{figure}[h!]
\centering
\includegraphics[width=80mm,height=70mm]{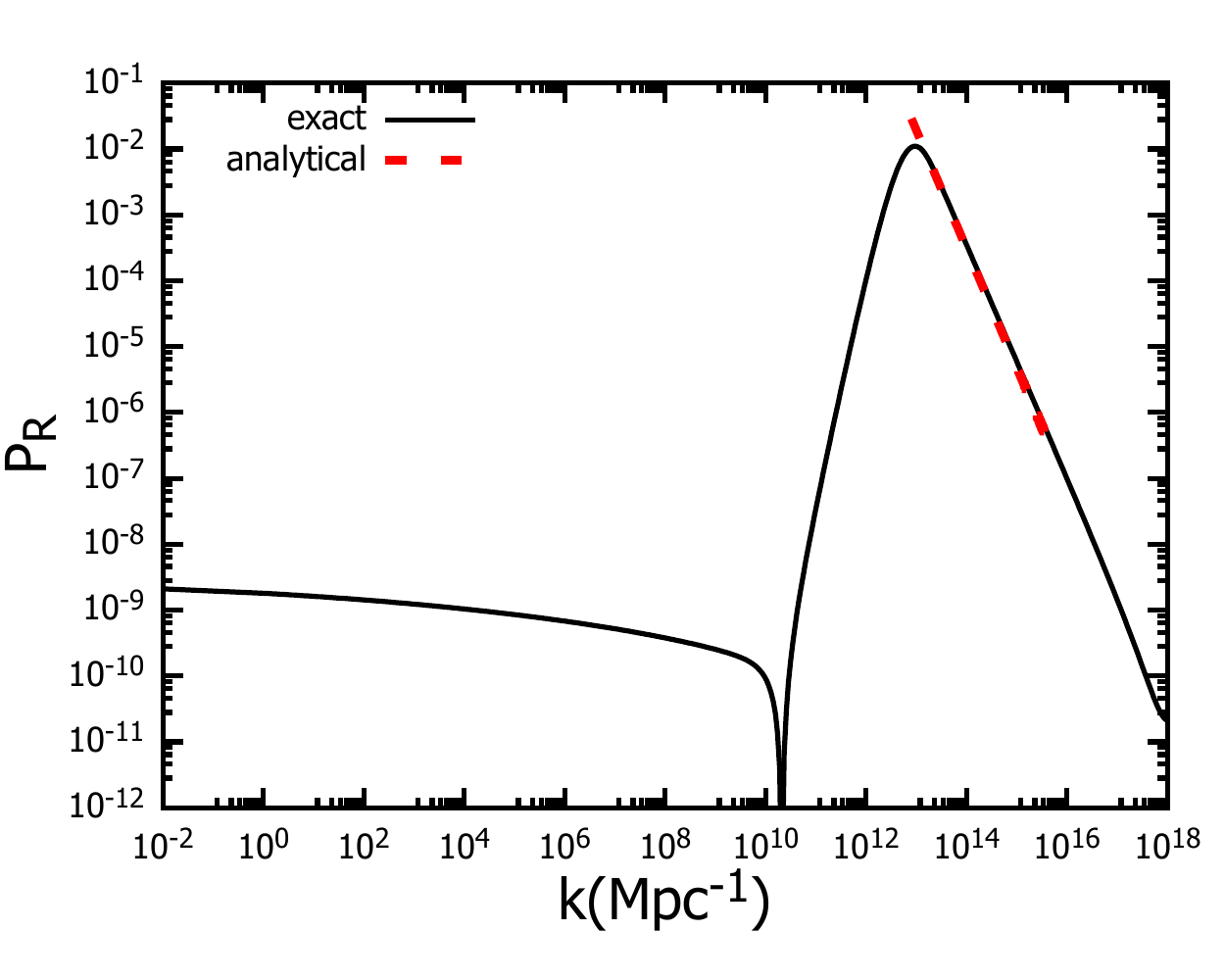}\includegraphics[width=80mm,height= 70mm]{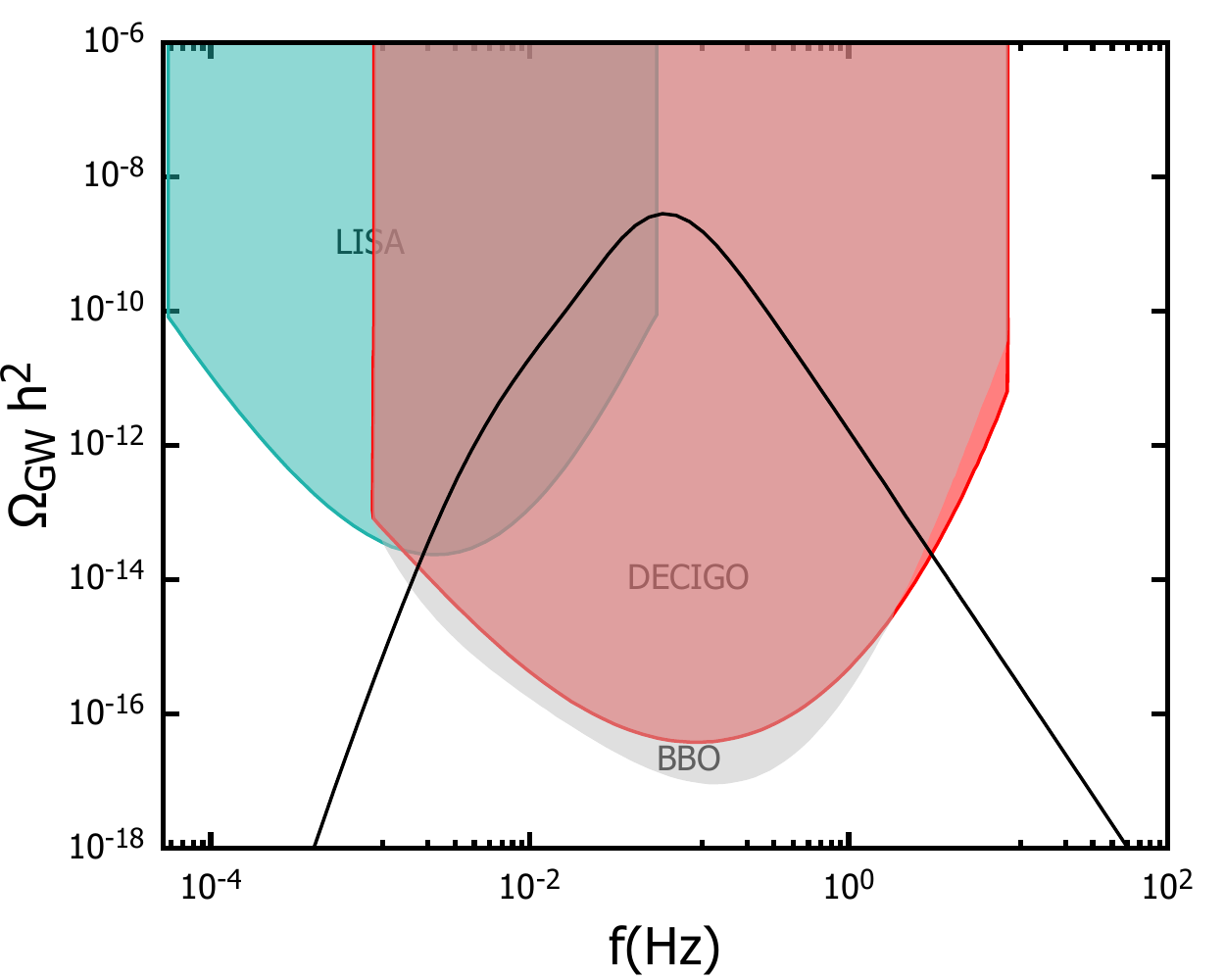}
\caption{\textcolor{black}{The power spectrum (left) with solid line (numerical) and with dashed (analytical from Eq.~\ref{eq:pr_analu}). The density of stochastic GWs (right). Both plots are for the case of modifying Wess-Zumino superpotential. We choose  $\lambda / \mu=1/3$, $c_1=1.9$, $c_2=-0.73$ and $b_1=3.9615$.}}
\label{f1}
\end{figure}

\begin{figure}[h!]
\centering
\includegraphics[width=80mm,height=70mm]{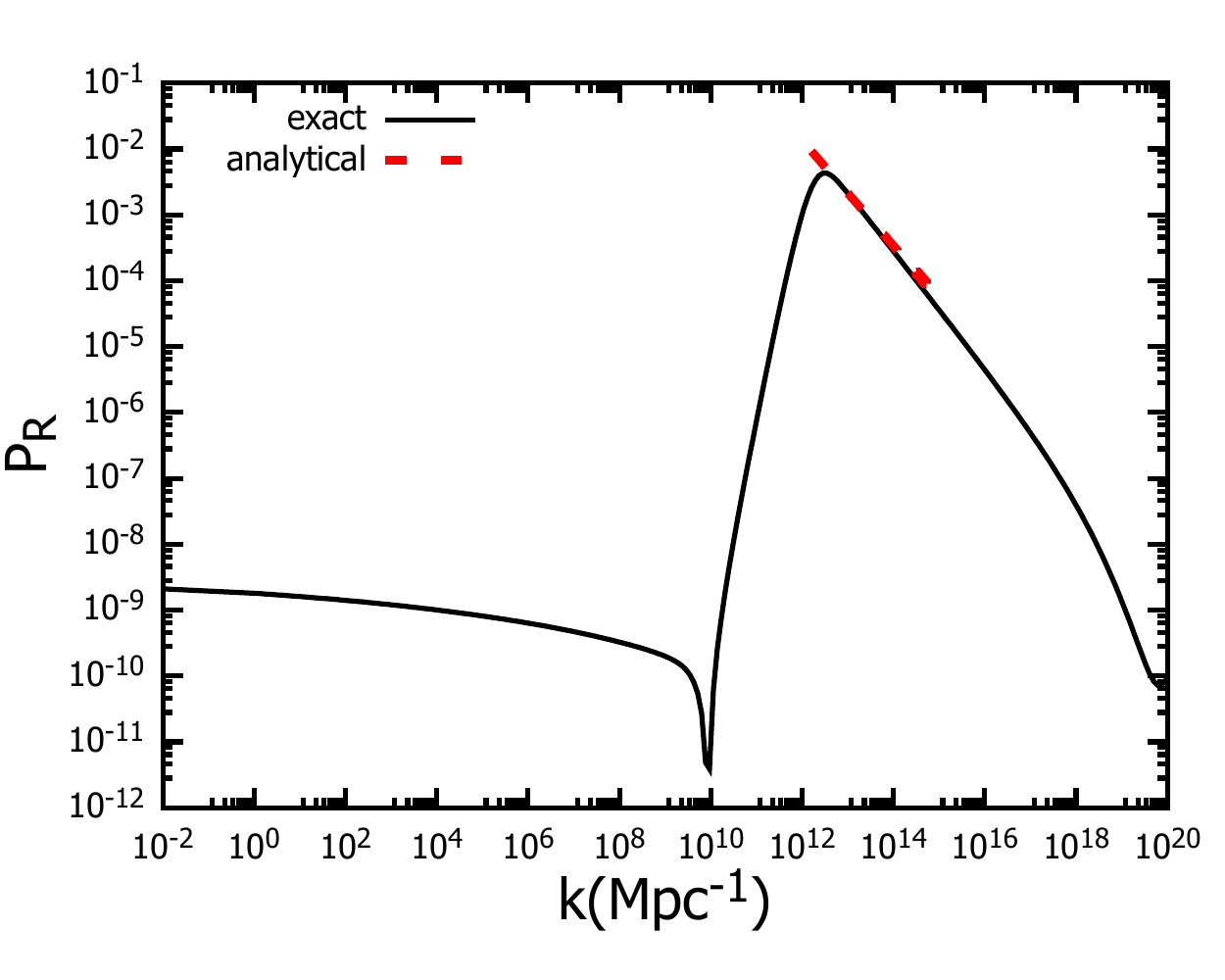}\includegraphics[width=80mm,height= 70mm]{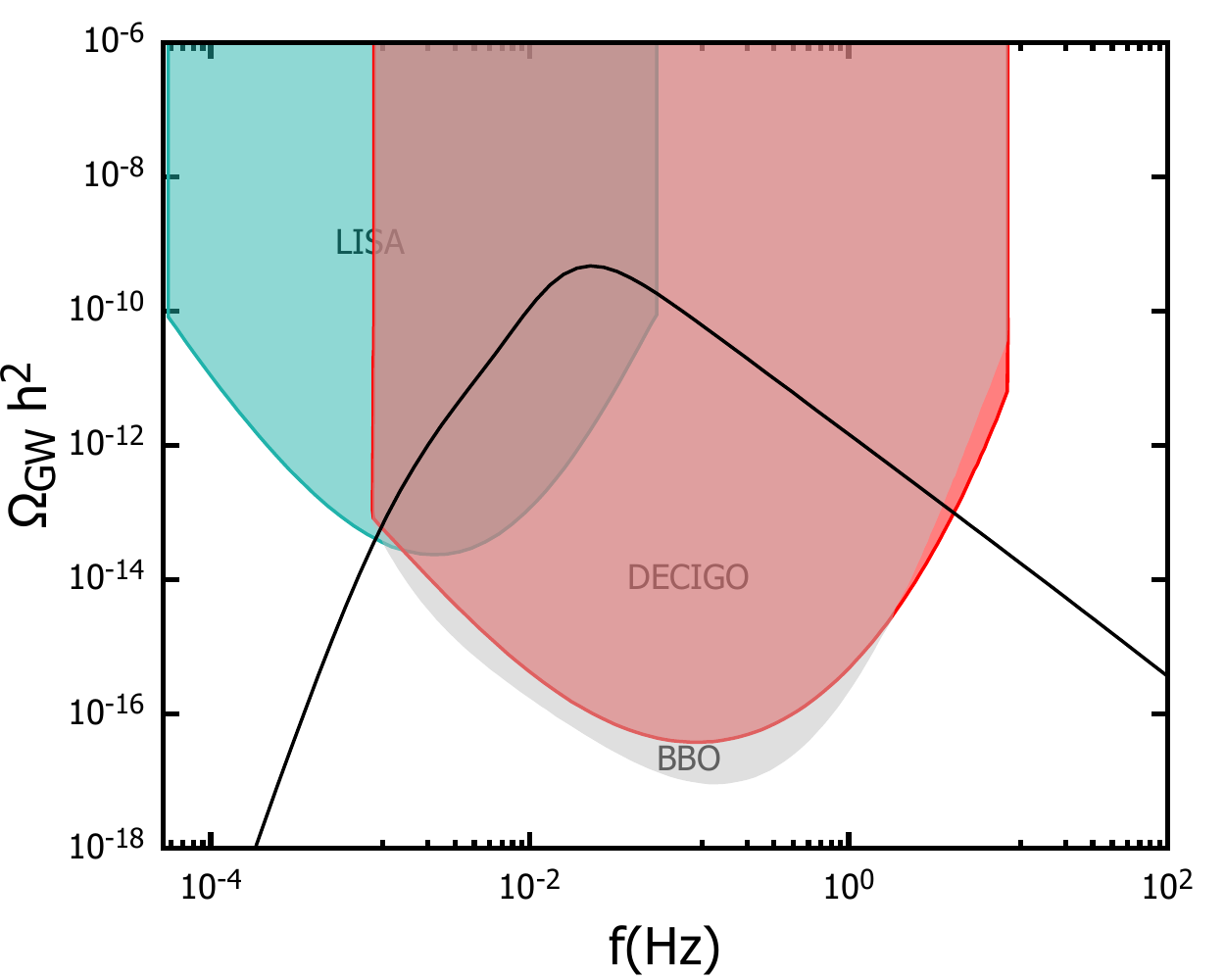}
\caption{\textcolor{black}{The power spectrum (left) with solid line (numerical) and with dashed (analytical from Eq.~\ref{eq:pr_analu}). The density of stochastic GW (right). Both plots are for the case of modifying Cecotti superpotential. We choose $c_3=13.2$, $l=1.00028$ and $b_2=8.575$.}}
\label{f2}
\end{figure}

Afterwards, we evaluate numerically the energy density of GWs, which is given in Eq.~(\ref{eq4.1}), as we described before. In Figs.~\ref{f1} and \ref{f2} we present the power spectrum and the density abundance of GWs for the cases (\ref{eq2.5}) and (\ref{eq2.6}) respectively. One can remark that the enhancement of scalar spectrum can be imprinted in the density of GWs. \textcolor{black}{We present  both the exact power spectrum given  from (\ref{eq1.6}) with solid black lines  and the spectrum from the power-law formula (\ref{eq:pr_analu})} with dashed red line. The choice of parameters, in order to achieve the inflection point in the effective scalar potential is given in the caption.
We notice that it is important 
the position of the peak,  to be within 
 $10^{12}-10^{14} Mpc^{-1}$,  in order to lie in the observational range of  the future GWs experiments.
 
\begin{figure}[h!]
\centering
\includegraphics[width=80mm,height=70mm]{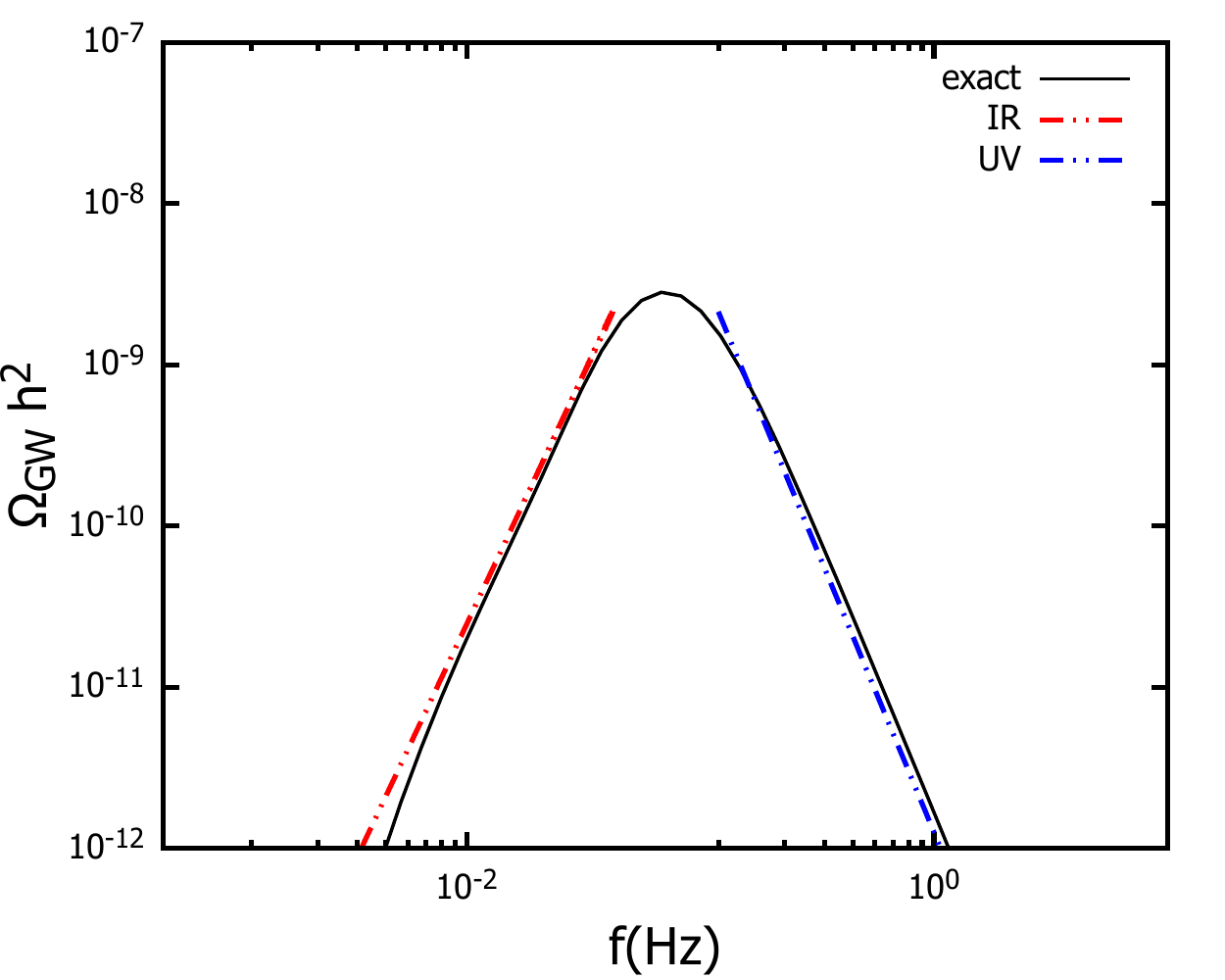}
\caption{\textcolor{black}{The energy density of GWs for the case  of modifying Wess-Zumino superpotential. Dashed lines corresponds to analytical approximation of the IR slope (red) and UV slope (blue).}}
\label{f11}
\end{figure}

 \textcolor{black}{In order  to  compare  with the numerical evaluation of the GWs abundance,   we show in Fig. \ref{f11} the corresponding example of Fig. \ref{f1}. We evaluate the IR slope given from Eq.(\ref{eq:gwIR})  with red dashed line and the UV slope given from Eq.(\ref{eq:gwUV}) with blue dashed line. We use  $n_{UV}=1.8$, that lies   in the region  $n_{UV}<4$.}

\subsection{Modifying the K\"ahler potential}
  \label{attr}
  
  In this section we present the results of the power spectrum and the abundances of GWs in the case of the   modified  K\"ahler potential, given in Eqs.~(\ref{eq2.8}) and (\ref{eq2.10}). This case is more labored than the previous one, because there is 
  analytic expression for 
  transformation of the  field,  in order to have canonical kinetic terms. Hence,   the Eq.~(\ref{eq12}) should be used. We proceed with the numerical integration of background and  curvature perturbations of the field,   assuming that we are in Bunch-Davies vacuum,  given in Eqs.~(\ref{eq21}), (\ref{eq1.2}) and (\ref{eq1.3}), as we have discussed  before.

\begin{figure}[h!]
\centering
\includegraphics[width=80mm,height=70mm]{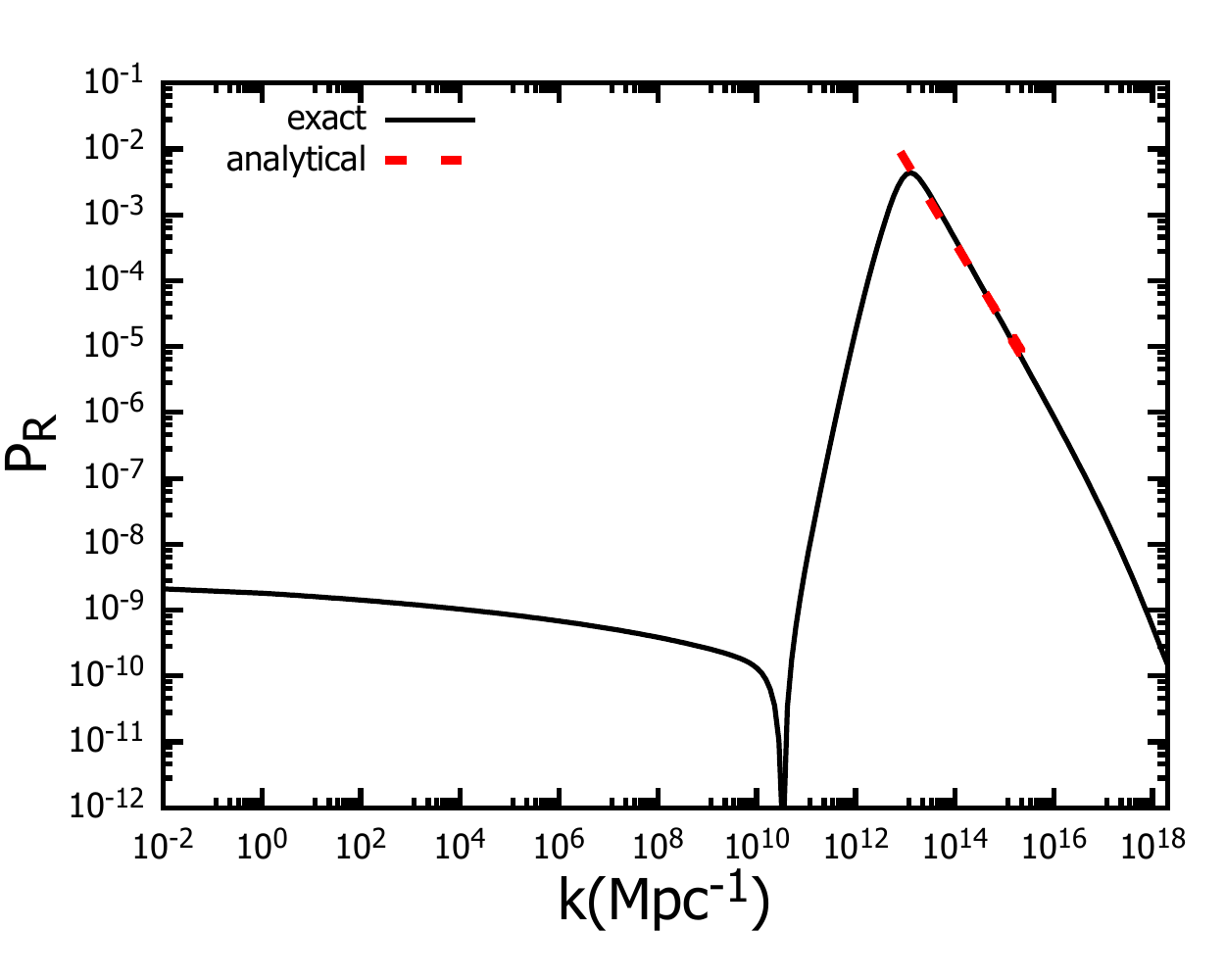}\includegraphics[width=80mm,height= 70mm]{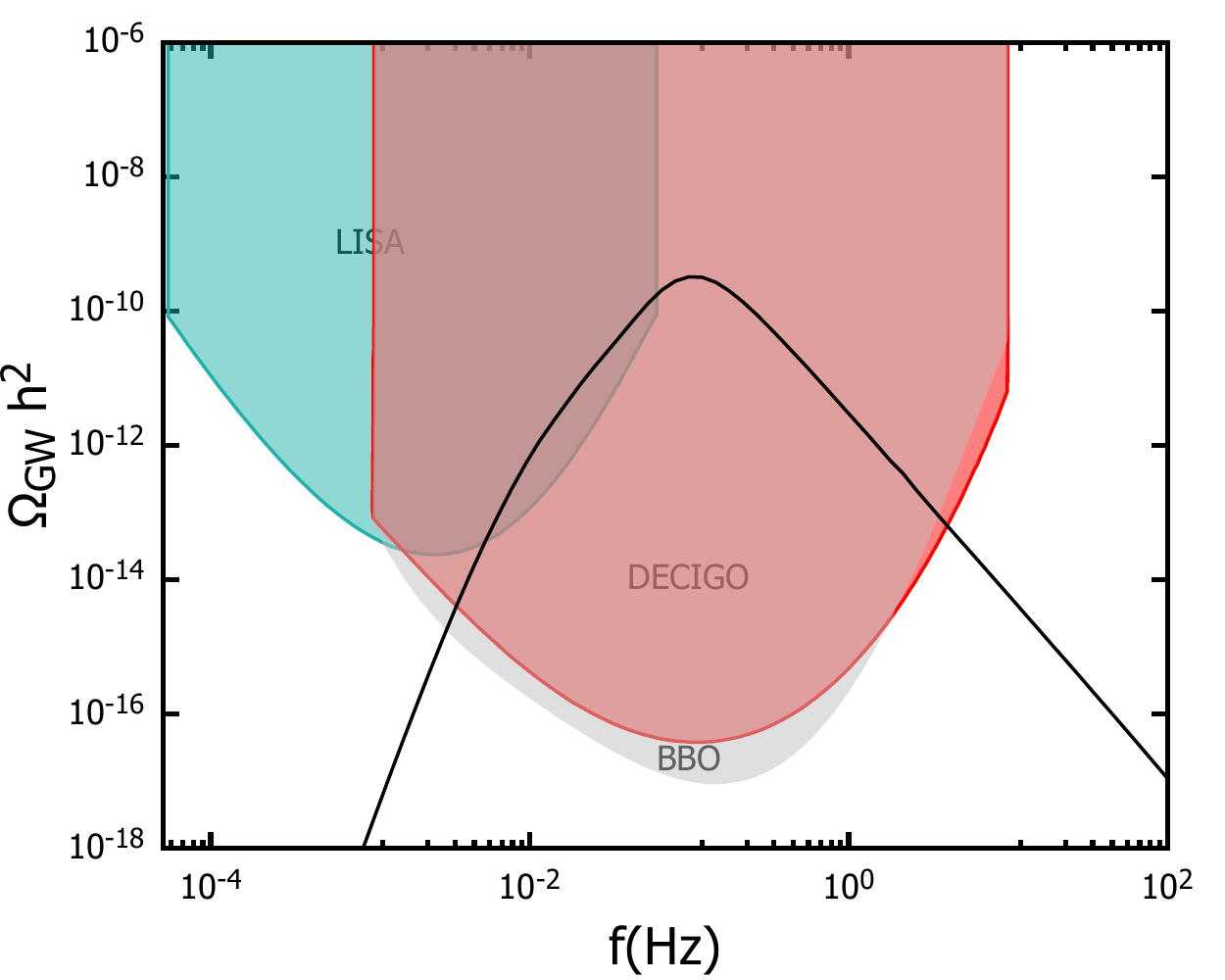}

\caption{\textcolor{black}{The power spectrum (left) with solid line (numerical) and with dashed (analytical from Eq.~\ref{eq:pr_analu}). The density of stochastic GWs (right) for the case of modifying Wess-Zumino. We choose for the parameters   $b_3=87.38$, $c=0.065$, $ c_3=-4$ and
$\lambda / \mu=0.33327$}}
\label{f3}
\end{figure}

\begin{figure}[h!]
\centering
\includegraphics[width=80mm,height=70mm]{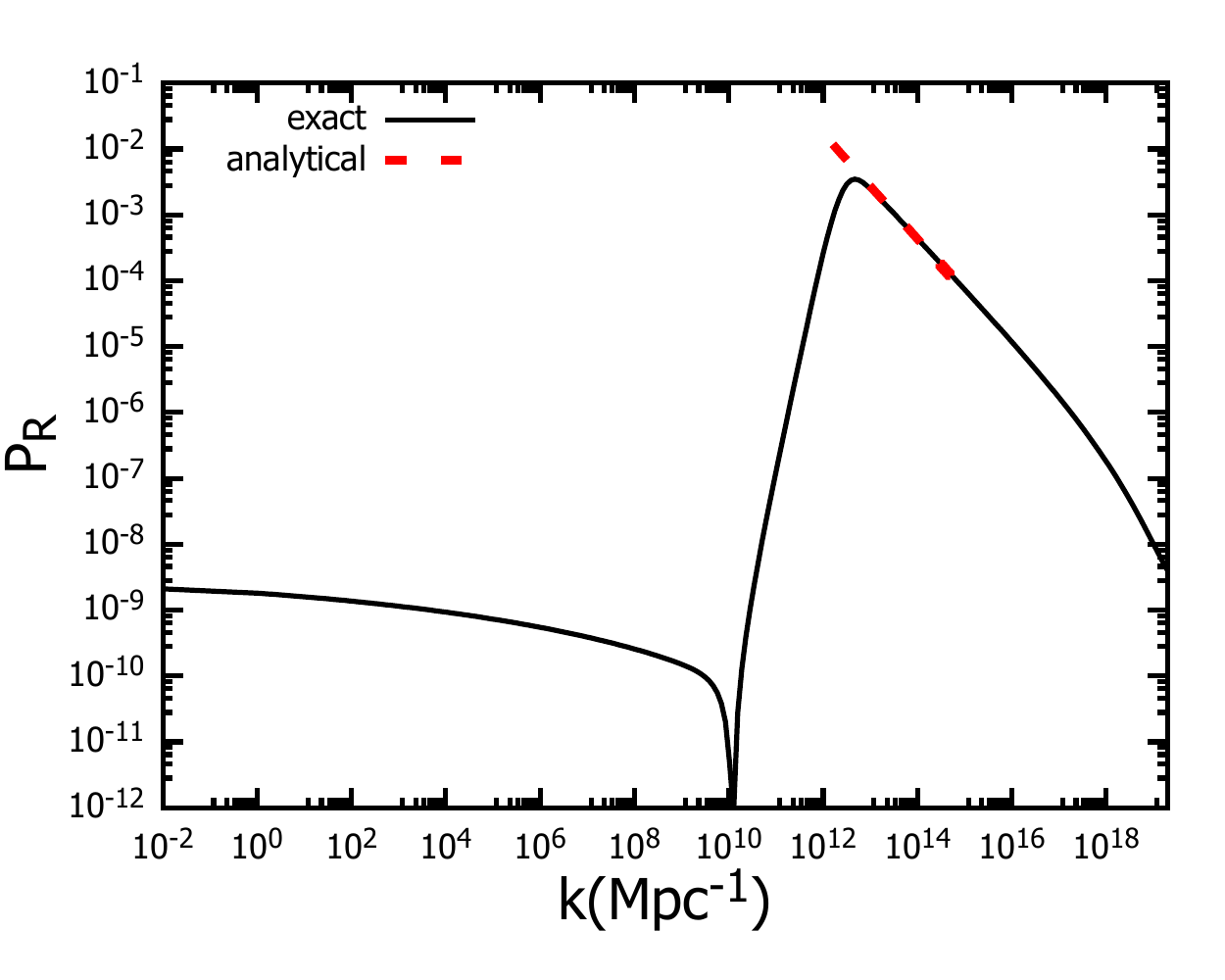}\includegraphics[width=80mm,height= 70mm]{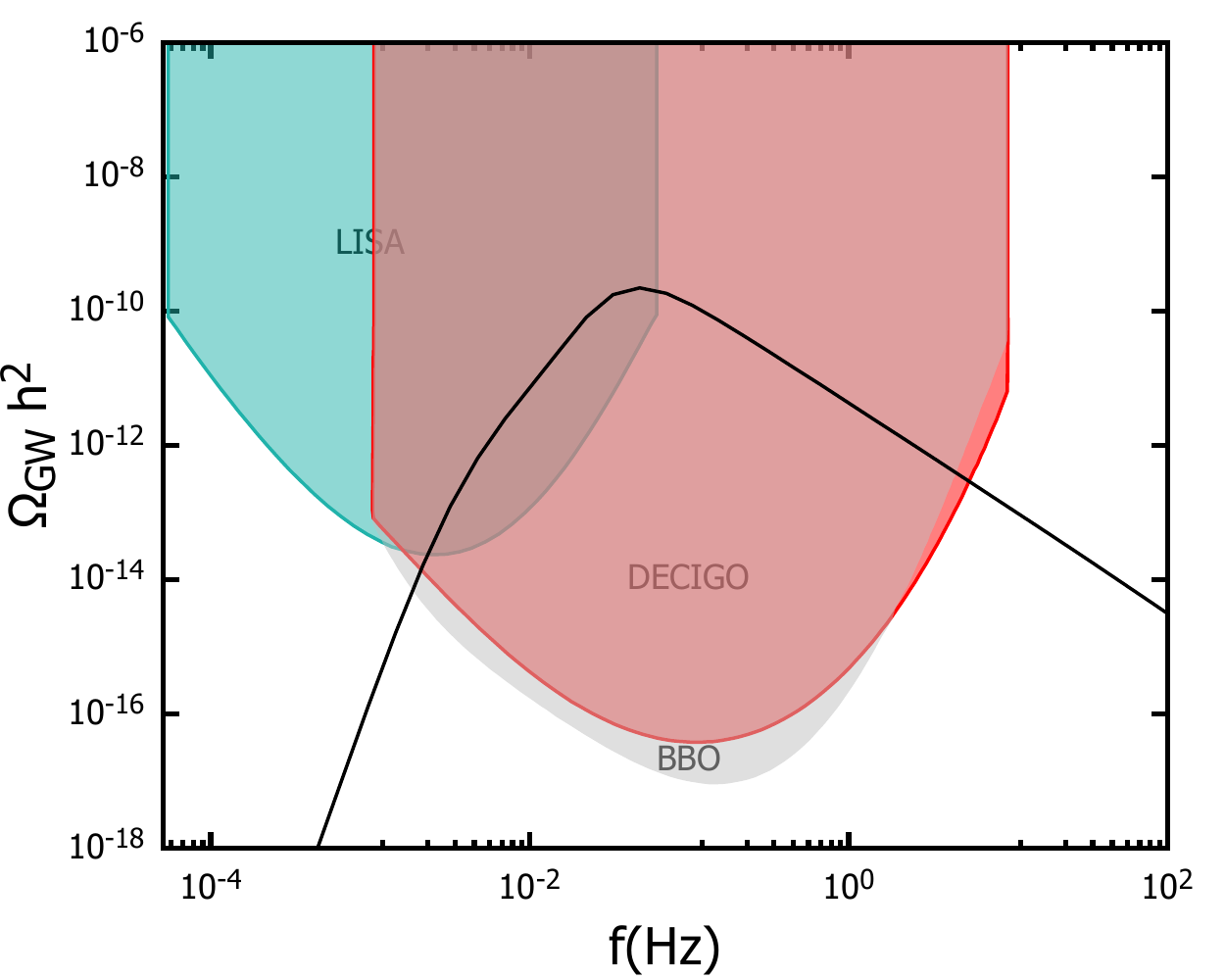}

\caption{\textcolor{black}{The power spectrum (left) with solid line (numerical) and with dashed (analytical from Eq.~\ref{eq:pr_analu}). The density of stochastic GWs (right) for the case of modifying Cecotti. The parameters   are  $ d=-0.054$,  $ b_4=7.51511$, $c_4=8.8$,
  $ \bar{\lambda}=-0.5$ and $\bar{\mu}=1.0$}}
\label{f4}
\end{figure}

In Figs.~\ref{f3} and \ref{f4} we present the resulting power spectrum and the density abundance of GWs for the Eqs.~(\ref{eq2.8}) and (\ref{eq2.10}),  respectively. The choice of initial conditions and the prediction for $n_s$ and $r$ are given in Table \ref{tabu2}. In this table we present   the initial condition for the canonical normalized field, which we denote with $\phi$. In  this Table we show that  the prediction of $n_s$ and $r$ are  consistent with the Planck constraints\cite{Ade:2015lrj,Akrami:2018odb}, as in the previous cases.

\begin{table}[h!]
\centering
 \begin{tabular}{|c|c|c|c|}
\hline
 & $\phi_{ic}$ & $n_s$ & $r$\\
\hline
1 & $4.899$ &$0.9612$ &$0.0121$  \\
\hline
2 & $4.097$ &$0.9601$ &$0.0092$\\
\hline

\end{tabular}
 \caption{Initial conditions for the field for the  Eqs. (\ref{eqwess}) and (\ref{eq2.8}) and the prediction of $n_s$ and $r$.}
 \label{tabu2}
\end{table}  

\textcolor{black} { In left panels in the Figs.~\ref{f3} and \ref{f4} we also show the numerical and analytical results of power spectrum from Eqs. (\ref{eq1.6}) and (\ref{eq:pr_analu}). The value of $n$ in Eq. (\ref{eq:pr_analu})  ranges from $0.15$ to $2.7$ \cite{Atal:2018neu}.  
In both models, either by modifying  the K\"ahler potential or  by modifying the superpotential), $n$ ranges from 0.8 to 1.8. Hence, in order to  estimate  the effect   of the 
non-gaussianities   the analysis of \cite{Atal:2018neu} can be used. 
In particular,  this related to the  peaks of power spectrum $P_{peak}=10^{-3}-10^{-2}$ as $f_{NL} \lessapprox 10^{-2}$. Generally, the amplitude $f_{NL}$ is defined  as  
\begin{equation}
    f_{NL}=\frac{5}{6}\frac{\mathcal{B}(k_1,k_2,k_3)}{P_R(k_1)P_R(k_2)+P_R(k_1)P_R(k_3)+P_R(k_2)P_R(k_3)} \,,
\end{equation}
where $\mathcal{B}$ is the bispectrum.
According to the study \cite{Atal:2018neu} the non-gausianities of models with an enhanced power spectrum can be large enough not to satisfy the constraints on the 
 PBHs abundance.
 We note that in this paper  we focus in  
 the production of GWs,  as the production  of the
 PBHs for these models has already presented elsewhere~\cite{Nanopoulos:2020nnh,Stamou:2021qdk}.}

The production of  PBHs using similar modifications in the SUGRA model has been studied in \cite{Stamou:2021qdk}.
 There,  it was  concluded that the maximum value of the peak of the power spectrum should be  around $3-5 \times10^{-2} $.
In this analysis here, that refers to the production of GWs this restriction does not apply. 
  
  That is, the  power spectra of GWs, will  be detected  by  future experiments, such as LISA or DECIGO 
  even if the high of the peak of the power spectrum is smaller than $10^{-2} $. \textcolor{black}{On the other hand, the abundance of the PBHs is exponentially sensitive
  to the mass fraction of the Universe that collapse to PBHs~ \cite{Harada:2013epa,Musco:2020jjb,Escriva:2020tak,Escriva:2019phb,Yoo:2018kvb}.} 
  Consequently, it is  expected that the amount of the  fine-tuning in 
the parameters of the extra terms, will be reduced. \textcolor{black}{ We  discuss this point  in  Appendix \ref{sect:app2}.}

\section{Conclusions}
In this paper we introduce four models in order to propose a possible scenario for the production of GWs,
that can be detected by future space-based experiments of GWs. 
\textcolor{black}{ A well-known way  to get amplified    production of the  GWs is an enhancement in the scalar power spectrum, since  the first and second order spectra are related. There are many mechanisms proposed in the literature,  in order to get  such an enhancement.
In  our study we employed  this of  the  near inflection point,  in the context 
of single field inflation. We embedded   this scenario in models  derived by no-scale SUGRA theory
and especially we consider models that  conserve  and break    the SU(2,1)/SU(2)$ \times$ U(1) symmetry.}

\textcolor{black}{First,  we modified  the  Wess-Zumino and the Cecotti superpotentials. These modifications leads to the  effective scalar potential, which is equivalent in both  bases $(T,\varphi)$ and $(y_1,y_2)$. Both of these models can conserve the  SU(2,1)/SU(2)$ \times$ U(1) symmetry, as we can  properly interplay between the fields  $y_1$ and $y_2$ and we can derive the same effective scalar potential. The same potential can be found, if we transform the fields and use the equivalent forms in the $(T,\varphi)$ basis. The equivalence between the models has been studied for the unmodified case. However, in our study we have the additional feature of an inflection point  that leads to a significant enhancement of the scalar power spectrum.  }

\textcolor{black}{ Secondly,  we keep these two superpotentials, Wess-Zumino and Cecotti,  unchanged  and we modify the K\"ahler potential in $(T,\varphi)$ basis. The resulting effective scalar potential can only be given numerically, as there is not an analytical solution with  canonical kinetic term.  Hence,  we derive  four inflationary potentials with a near inflection point.
}

This enhancement of scalar power spectrum is expected to be imprinted in the energy density of GWs.  In order to evaluate this energy density, we calculate numerically the perturbations of the field for all the proposed potentials. \textcolor{black}{We evaluate the scalar power spectrum and then the corresponding energy densities of GWs.}  We show that these models give sizable GWs spectra,  which  can be detected by  the future experiments. \textcolor{black}{Last but not least, 
all the  models presented in this work are consistent with 
the observational  constraints on inflation.}


\appendix

\section{\textcolor{black}{The  forms of the scalar potential}}
\label{sect:app1}
\textcolor{black}{
In general  the scalar potential can be evaluated in the framework of supergravity, if we use the following expression:
\begin{equation}
V= e^{K} (D_{\Phi}WK^{\bar{\Phi} \Phi} D_{\bar{\Phi}}\bar{W} - 3|W|^2) \, , 
\label{eq:app:pot}
\end{equation}
 where
\noindent
\begin{equation}
D_{\Phi}W=\frac{\partial W}{\partial \Phi}+ \frac{\partial K}{\partial \Phi}W .
\end{equation}
In this Appendix we sum up the four models proposed in the text and we present  the effective scalar potentials for the two first models.}

\subsection{\textcolor{black}{Model I}}
\textcolor{black} {For the choice of K\"ahler potential
\begin{equation}
K=-3\ln(1-\frac{|y_1|^2}{3}-\frac{|y_2|^2}{3})
\label{eq:app:ky1y2}
\end{equation}
we modify the superpotential given in Eq. (\ref{eq6}) as: 
\begin{equation}
W=\left[ \frac{\hat\mu}{2} \left(y_1^2 +\frac{y_2^2 y_1}{\sqrt{3}}\right) -\lambda \frac{y_1^3}{3}\right]  
\left[ 1+e^{-b_1y_1^2} \left(c_1 {y_1}^2+c_2{y_1}^4 \right)\right] \, 
\end{equation}
where $\hat{\mu}$, $\lambda$, $b_1$, $c_1$ and $c_2$ are free parameters.
If we use Eq. (\ref{eq:app:pot}), the effective scalar potential becomes:
\begin{equation} 
\begin{split}
V/ \mu^2&=\sinh ^2\left(\frac{\phi}{\sqrt{6}}\right) \Big[ 9 \tanh ^4\left(\frac{\phi}{\sqrt{6}}\right) e^{-6 b_1 \tanh ^2\left(\frac{\phi}{\sqrt{6}}\right)} \left(\mu -2 \lambda  \tanh \left(\frac{\phi}{\sqrt{6}}\right)\right)^2\times\\&
\times\left(\tanh ^2\left(\frac{\phi}{\sqrt{6}}\right) (6 {c_2}-3 b_1 c_1)-9 b_1 {c_2} \tanh ^4\left(\frac{\phi}{\sqrt{6}}\right)+c_1 \right)^2- \\&-3 \sinh ^2\left(\frac{\phi}{\sqrt{6}}\right) e^{-6 b_1 \tanh ^2\left(\frac{\phi}{\sqrt{6}}\right)} \left(\mu -2 \lambda  \tanh \left(\frac{\phi}{\sqrt{6}}\right)\right) \times\\& \times \left(2 \mu -6 \lambda  \tanh
   \left(\frac{\phi}{\sqrt{6}}\right)+\mu  \tanh ^2\left(\frac{\phi}{\sqrt{6}}\right)\right)\cosh ^2\left(\frac{\phi}{\sqrt{6}}\right) \left(\mu -3 \lambda  \tanh \left(\frac{\phi}{\sqrt{6}}\right)\right)^2 \\& \times  \left(3 \tanh ^2\left(\frac{\phi}{\sqrt{6}}\right) e^{-3 b_1 \tanh
   ^2\left(\frac{\phi}{\sqrt{6}}\right)} \left(c_1+3 {c_2} \tanh ^2\left(\frac{\phi}{\sqrt{6}}\right)\right)+1\right)^2+  \\& + \left(e^{3 b_1 \tanh ^2\left(\frac{\phi}{\sqrt{6}}\right)}+3 c_1 \tanh
   ^2\left(\frac{\phi}{\sqrt{6}}\right)+9 {c_2} \tanh ^4\left(\frac{\phi}{\sqrt{6}}\right)\right)\times \\& \times  \left(3 \tanh ^2\left(\frac{\phi}{\sqrt{6}}\right) (b_1 c_1-2 {c_2})+9 b_1 {c_2} \tanh
   ^4\left(\frac{\phi}{\sqrt{6}}\right)-c_1\right)\Big]
\end{split}
\end{equation}
where $\hat \mu = \mu \sqrt{c/3}$.
We have  considered  that the inflationary direction for this example is given as:
\begin{equation}
y_1=\varphi,\quad y_2=0.
\label{eq:app:direction}
\end{equation}
Moreover, in order to have fix the non- canonical kinetic term we consider the redefinition of the field: 
\begin{equation}
y_1=-\sqrt{3}\tanh\left( \frac{\phi}{\sqrt{6}}\right) \,. 
\end{equation}
We remark that the same effective scalar potential can be derived if we have the following transformations: $y_1 \rightarrow -y_1$ and $y_2 \rightarrow -y_2$ \cite{Stamou:2021qdk,Ellis:2018zya}.}
 
\textcolor{black}{The same form of the potential can be derived, if we use the basis $(T,\varphi)$ with the tranformation described in the text from Eqs.(\ref{eq2.3}), (\ref{k2(3)}) and (\ref{eq2.4}). Hence, we have the following forms:
\begin{equation}
K=-3 \ln(T +\bar{T}- \frac{|\varphi|^2}{3})
\label{eq:app:ktphi}
\end{equation}}

\textcolor{black}{\begin{equation}
 W=\left( \frac{\hat{\mu}}{2}\varphi^2 -\frac{\lambda}{3} \varphi^3\right)\left(  1+ e^{-b_1 \left( \frac{2 \varphi}{1+ 2 T} \right)^2} \left[ c_1 \left( \frac{2 \varphi}{1+ 2 T} \right)^2+ c_2 \left( \frac{2 \varphi}{1+ 2 T} \right)^4\right] \right) .
\end{equation}
We choose the following direction $T=Im\varphi=0$ and $Re\varphi=\phi$ and we apply  Eq.~(\ref{eq:app:pot}). }

\subsection{\textcolor{black}{Model II}}
\textcolor{black}{For the choice of K\"ahler potential (\ref{eq:app:ky1y2})
we modify the superpotential given in Eq. (\ref{eq9}) as: 
\begin{equation}
 W=m\left(-y_1y_2 
  +\frac{y_2y_1^2}{l\sqrt{3}}\right)\left(1+c_3e^{-b_2{y_1}^2}{y_1}^2\right) \, .
\end{equation}
From Eq. (\ref{eq:app:pot}) we derive the form of effective scalar potential, which is given as follows:
\begin{equation}
\begin{split}
V/m^2&=\frac{3}{l^2}  \sinh ^2\left(\frac{\phi}{\sqrt{6}}\right) \cosh ^2\left(\frac{\phi}{\sqrt{6}}\right) e^{-6 b \tanh ^2\left(\frac{\phi}{\sqrt{6}}\right)} \left(l-\tanh
   \left(\frac{\phi}{\sqrt{6}}\right)\right)^2\\&\times \left(e^{3 b \tanh ^2\left(\frac{\phi}{\sqrt{6}}\right)}+3 c_3 \tanh ^2\left(\frac{\phi}{\sqrt{6}}\right)\right)^2.
\end{split}
\end{equation}
As in model 1, we choose the inflationary direction (\ref{eq:app:direction}) and we fix the non- canonical kinetic terms. This potential has the following free parameters: $l$, $b_2$ and $c_3$. }

\textcolor{black}{We can use the other form of the K\"ahler potential (\ref{eq:app:ktphi}) and with proper  tranformations given from Eqs.(\ref{eq2.3}), (\ref{k2(3)}) and (\ref{eq2.4}) we can take the equivalent form of the superpotential in the basis $(T,\varphi)$.  The superpotential is given as follows:
\begin{equation}
W= \frac{\sqrt{3}}{2} m \varphi \Big( \frac{1}{2} -T\Big) \left( -1-2T -\frac{1-2T}{l} \right) \left[ 1+ 3c_3 e^{-3b_2 \left( \frac{1-2T}{1+2T} \right)^2}\left( \frac{1-2T}{1+2T} \right)^2\right] \, .
\end{equation}}

\subsection{\textcolor{black}{Model III}}
\textcolor{black}{In this model we do not change the superpotential but the K\"ahler potential in $(T,\varphi)$ basis. For superpotential we choose this of Wess Zumino model described in Eq.~(\ref{eq6}). 
\begin{equation}
K=-3 \ln( T+ \bar{T} - \frac{\varphi \bar{\varphi}}{3}+ c_4e^{-b_3(\varphi +\bar{\varphi})^2}(\varphi +\bar{\varphi})^4) \,,
\end{equation} 
\begin{equation}
W= \frac{\hat \mu}{2} \varphi^2  - \frac{\lambda}{3} \varphi^3 .
\end{equation} 
We can evaluate the scalar potential  from (\ref{eq:app:pot}). In this case the non-canonical kinetic term can be fixed only numerically.  The direction of inflation is $T=c/2$ and $\varphi=\phi$. The free parameters for this case are $c$, $\lambda$, $\mu$, $c_3$,  $b_3$ and $c$. }

\subsection{\textcolor{black}{Model IV}}
\textcolor{black}{In this model we modify the  described in Eq.(\ref{eq:app:ktphi})
and we keep unchanged the Cecotti superpotential given in (\ref{eq9}). Hence we have:
\begin{equation}
K=-3 \ln( T+ \bar{T} - \frac{\varphi \bar{\varphi}}{3}+ F( T+ \bar{T},\varphi+ \bar{\varphi}))
\label{eq:app2.8}
\end{equation} 
\begin{equation}
W=\sqrt{3}m \varphi\left(T- \frac{1}{2}\right)
\label{eq:app2.9}
\end{equation}
\noindent
where $F( T+ \bar{T},\varphi+ \bar{\varphi})$ is a function of both chiral fields. We choose the following form
\begin{equation}
\label{eq:app2.10}
 F( T+ \bar{T},\varphi+ \bar{\varphi})=e^{-d (T+ \bar{T})}\left(c_4 e^{-b_4(T+ \bar{T})}(T+ \bar{T})^2 +\bar{\lambda}(\varphi+ \bar{\varphi}+ \bar{\mu})\right) \, ,
\end{equation}
and we assume that the inflationary direction is
\begin{equation}
T=ReT=\phi, \quad ImT=\varphi=0. 
\end{equation}
The scalar power spectrum can be evaluated numerically by using  Eq. (\ref{eq:app:pot}). We need to consider the transformation of the field in order to have canonical kinetic term. This redefinition can be evaluated only numerically as in the previous case. The free parameters are $m$, $d$, $b_4$, $c_4$, $\bar{\lambda}$ and $\bar{\mu}$.}

\section{\textcolor{black}{Fine-tuning  analysis and a comparison between the models}}
\label{sect:app2}
It is well-known that the enhancement of scalar power spectrum, which occurs due to the inflection point in the effective scalar potential requires a lot of fine-tuning~\cite{Hertzberg:2017dkh}.  
The value of this enhancement can be smaller in the case of studying  the production of GWs than in the  case of studying the amount of DM from the PBHs.   In Ref.~\cite{Stamou:2021qdk} there is a discussion explaining how the parameters of the potential presented in this work arise. In this Appendix we analyse the level of fine-tuning by considering the parameter $b_i$, which is presented throughout this work and   is the parameter,  which depends on the power spectrum's peak and demands more fine-tuning.

The role of the parameter $b_i$ is to trigger  an enhancement in the power spectrum. In order to analyze the level of fine-tuning, we calculate the parameter $\Delta_b$, which is shown in Refs.~\cite{Barbieri:1987fn,Leggett:2014mza} and it is given as the max value of the follow quantity
\begin{equation}
\Delta_b=\left| \frac{\partial \ln(P_R^{peak})}{\partial \ln(b_i)} \right| \,  . 
\end{equation}
In the following, we study  the fine-tuning of the function $P_R^{peak} (b_i)$. Large value of the maximum of the quantity $\Delta_b$ means that a high level of fine-tuning is required. As we mentioned before, we expect that the fine-tuning of the parameters can be decreased in the study of generation of GWs, due to its wider range of peak's height in comparison with the production of PBHs.

\begin{figure}[h!]
\centering
\includegraphics[width=95mm,height=70mm]{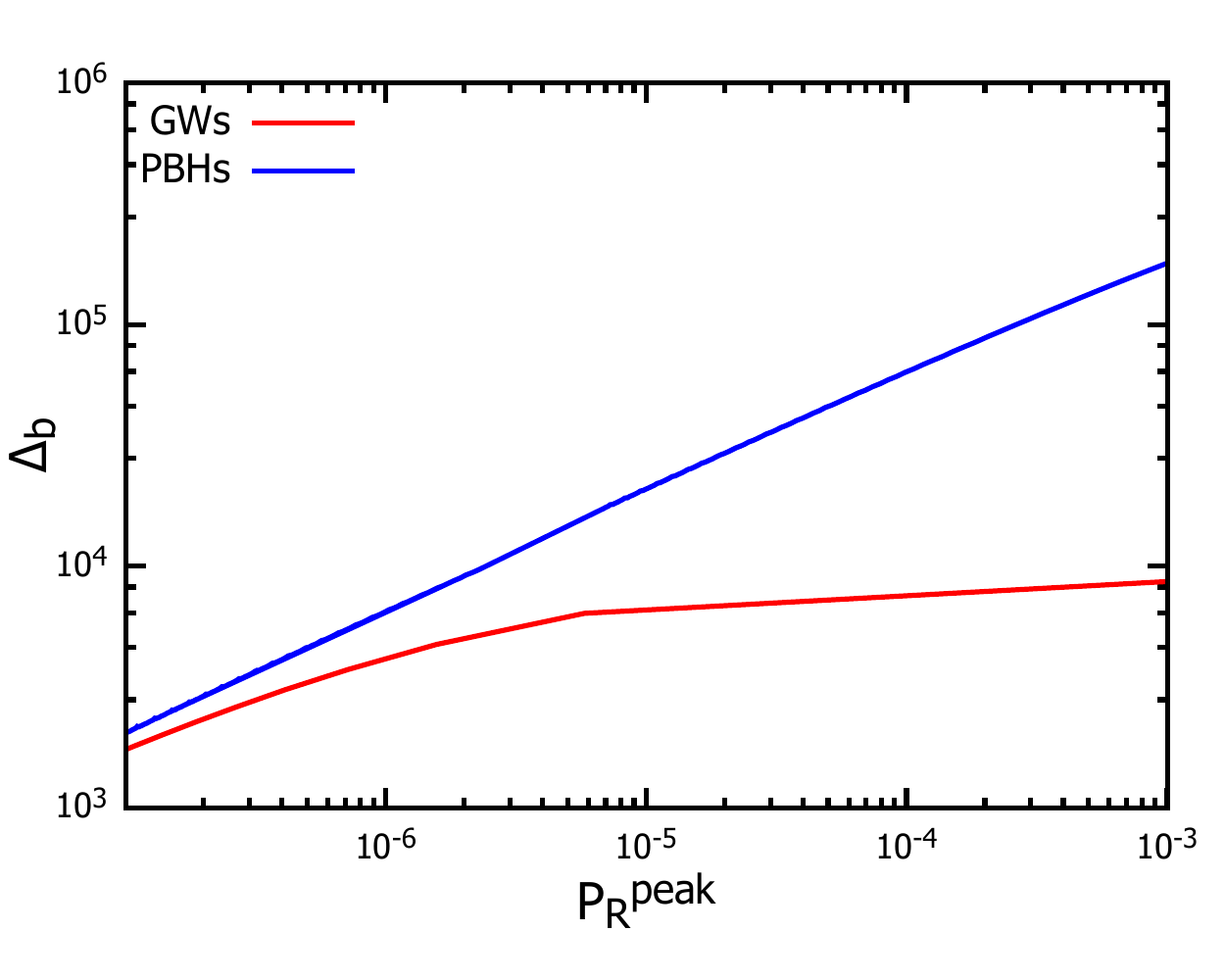}

\caption{\textcolor{black}{The level of fine-tuning for the case (\ref{eqwess}).} }
\label{f5}
\end{figure}

In Fig.\ref{f5} we show the quantity $\Delta_b$ as a function of the peak of the power spectrum. \textcolor{black}{ In particular we show how the quantity $\Delta_b$ varies in respect to  the values of the spectrum's peak in the case of study PBHs  (blue line) and in the case of GWs (red line). The case for this plot is  the model  described from the Eqs.(\ref{eqwess}) and( \ref{equation1}) (model III of Appendix \ref{sect:app1}). As one can notice we need more fine-tuning when we study PBHs. This occurs not only due to smaller peaks, which are needed in the study of GWs, but also due to the specific value of the threshold presented in the evaluation of the production of  PBHs. Specifically, in order to obtain a significant amount of DM from PBHs and to use the proposed values for threshold, as analysed in \cite{Harada:2013epa,Musco:2020jjb,Escriva:2020tak,Escriva:2019phb,Yoo:2018kvb}, we need to properly adjust the extra parameters  of the model. This increases the fine-tuning in the study of PBHs. }
 
We remark that in the previous study of Refs~\cite{Stamou:2021qdk}, where the fine-tuning for the PBHs production  is analysed, the   maximum value of $(\Delta_b)$ was at the order of magnitude of $10^{6}$. In a relevant study of  Ref. ~\cite{Hertzberg:2017dkh} the amount of fine-tuning for PBHs was at level of  $10^{8}$.  
It is possible that the fine-tuning can be decreased further.
In Table \ref{tabu4}, we show the value of the maximum  of  $\Delta_b$, assuming that we have the minimal enhancement of peak of the power spectrum, in order to predict the future space based experiment DECIGO. In other words,  we calculate the max( $\Delta_b$) by considering that the peak of power spectrum is around $k =10^{-5}$. In  Table~\ref{tabu4} we present, for comparison, the maximum value of $\Delta_b$ for PBHs, as adopted in  Refs.~\cite{Nanopoulos:2020nnh,Stamou:2021qdk}. 
One can notice that there is sizable  difference with respect to  the case of the production  of the GWs.
Specifically, the fine-tuning in our current analysis is smaller    at least two orders of magnitude.
\begin{table}[h!]
\centering
 \begin{tabular}{|c|c|c|}
\hline
  case& max$(\Delta_b)_{PBHs}$ & max$(\Delta_b)_{GWs}$\\
\hline
\hline
 1& $7.9$ $\times10^5$&$3.5$ $\times10^3$  \\
\hline
  2 & $ 9.8$ $\times10^5$ &$ 3.6$ $\times10^3$\\
\hline
 3& $1.6$ $\times10^5$ &{$8.5\times10^2$}  \\
\hline
 4& $4.2$ $\times10^6$&$5.6$ $\times10^4$\\
\hline
\end{tabular}
 \caption{The maximum  value for  $\Delta_b$ for the cases in  Eqs.~(\ref{eq2.5}), (\ref{eq2.6}),(\ref{eqwess}) and (\ref{eq2.8}) respectively. The abundances of PBHs are adopted by \cite{Nanopoulos:2020nnh,Stamou:2021qdk}. Details given are in main text. }
 \label{tabu4}
\end{table} 

\begin{figure}[h!]
\centering
\includegraphics[width=78mm]{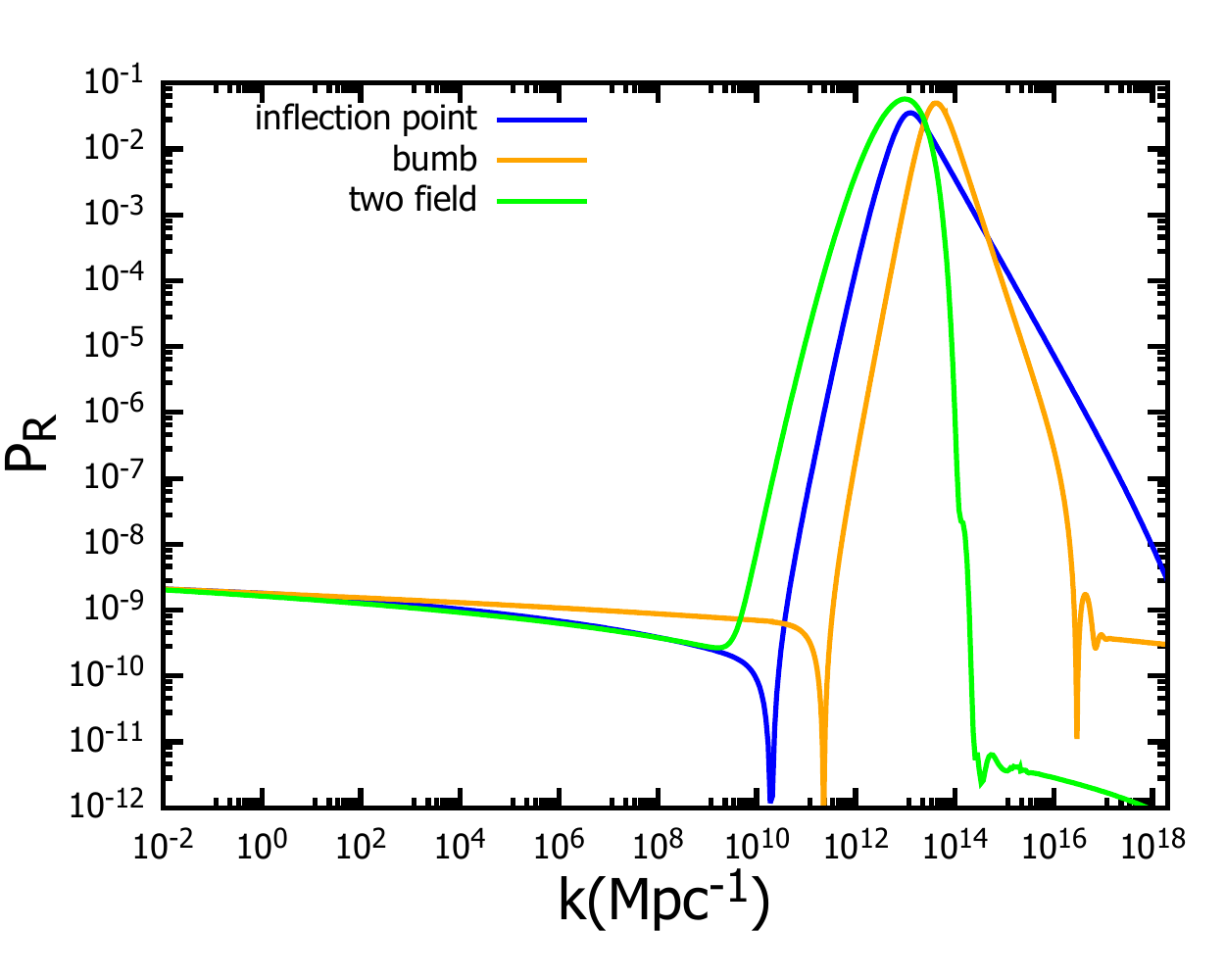}
\includegraphics[width=78mm]{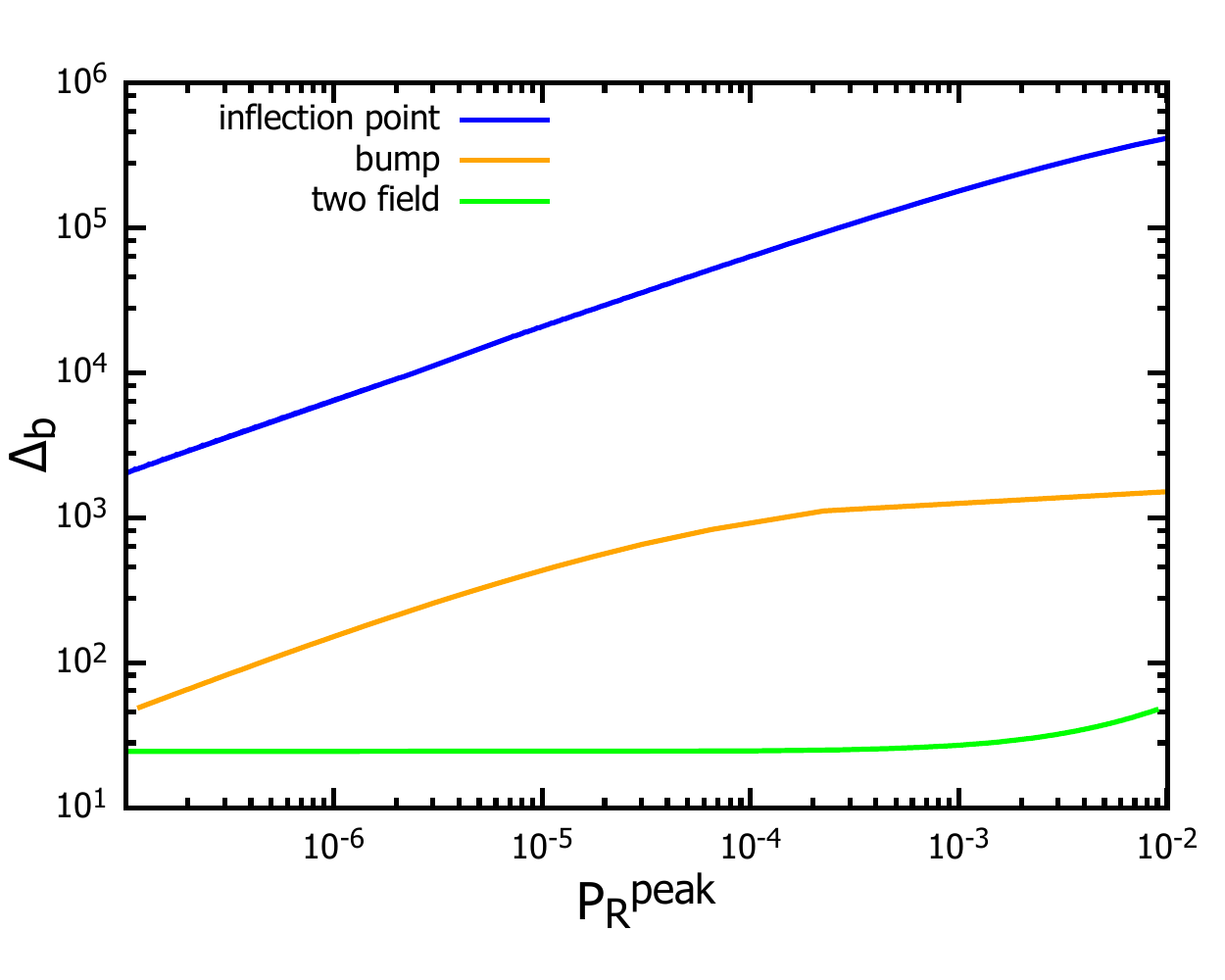}

\caption{\textcolor{black}{The level of fine-tuning for our model (blue) with comparison with this amount from other models.}}
\label{f7}
\end{figure}

\textcolor{black}{Finally, in order to have a comparison with the previous mechanisms  with an enhancement in scalar power spectrum analysed in the literature, we depict the quantity $\Delta_b$ for different models. In particular,  in Fig.\ref{f7}  we show the power spectrum for the case given Eqs.(\ref{eqwess}) and( \ref{equation1}) with blue lines. Orange lines corresponds to the model presented in \cite{Zheng:2021vda} where the enhancement of power spectrum comes from a bulky potential. Green lines to the two- model shown in \cite{Braglia:2020eai}. In this work a two field model with a non-canonical kinetic term has been proposed.   In both these models the underlying parameter b needs fine-tuning in order to obtain the proper enhancement. The left panel of the Fig.\ref{f7}  shows the power spectrum of these models, reproduced by the methodology of subsection \ref{Curvature Perturbations} and the right panel shows the quantity  $\Delta_b$. We  notice that models with an inflection point have more fine-tuning than the other ones. Models with a bumb feature in the potential have significantly  less  fine-tuning   than the models with an inflection point. Finally, models with two fields experience the less fine-tuning of all cases.   }

\vspace{1cm}
\section*{Acknowledgments}
 This research work was supported by the Hellenic Foundation for Research and Innovation (H.F.R.I.) under the ``First Call for H.F.R.I. Research Projects to support Faculty members and Researchers and the procurement of high-cost research equipment grant'' (Project Number: 824).

\bibliographystyle{elsarticle-num}

\end{document}